\documentclass[twocolumn]{aastex6}
\usepackage{graphicx}
\usepackage{amsmath}
\usepackage{subfigure}

\usepackage{rotating,natbib}
\usepackage{epstopdf}
\newcommand{\kms}{km s$^{-1}$}
\newcommand{\kmsMpc}{km s$^{-1}$ Mpc$^{-1}$}
\newcommand{\Ho}{$H_0$}

\slugcomment{Submitted to ApJ}
\shorttitle{Action Dynamics of the Local Supercluster}
\shortauthors{Shaya, Tully, Hoffman \& Pomar\`ede}
\begin{document}
\received{2017, August 16}

\title{Action Dynamics of the Local Supercluster}
\author{Edward J. Shaya}
\affil{Astronomy Department, University of Maryland,
    College Park, MD 20742, USA}
\email{eshaya@umd.edu}
\author{R. Brent Tully}
\affil{Institute for Astronomy, University of Hawaii, Honolulu, HI 96822, USA}
\and
\author{Yehuda Hoffman}
\affil{Racah Institute of Physics, Hebrew University, Jerusalem 91904, Israel}
\and
\author{Daniel Pomar\`ede}
\affil{Institut de Recherche sur les Lois Fondamentales de l'Univers, CEA, Universit\'e Paris-Saclay, F-91191 Gif-sur-Yvette, France}

\begin{abstract} 
The fully nonlinear gravitationally induced trajectories of a nearly complete set of galaxies, groups, and clusters in the Local Supercluster are constructed in a Numerical Action Method (NAM) model constrained by data from the \textit{CosmicFlows} survey and various distance indicators. We add the gravity field due to inhomogeneities external to the sample sphere by making use of larger scale peculiar flow measurements. Assignments of total masses were made to find the best overall set of mutual attractions, as determined by a goodness criterion based on present day radial velocities, individually for the Virgo Cluster, M31, and the Milky Way (MW), and via a mass-to-light ratio relationship for other masses. The low median chi-square found indicates the model fits the present day velocity flow well, but a slightly high mean chi-square may indicate that some masses underwent complex orbits. The best fit, when setting the value of $H_0$ to the \textit{CosmicFlows} value of 75 km/s/Mpc and the WMAP value for $\Omega_m=0.244$ consistent with that $H_0$, occurs with the following parameters: $\Omega_{orphan} = 0.077\pm0.016$, $M/L_K = 40 \pm 2 L_{10}^{0.15} M_\odot/L_\odot$ ($L_{10}$ is K-band luminosity in units of $10^{10} L_\odot$), a Virgo mass of $6.3\pm0.8 \times 10^{14} M_\odot$ ($M/L_K = 113\pm15 M_\odot/L_\odot$), and a mass for the MW plus M31 of $5.15\pm0.35 \times 10^{12} M_\odot$. The best constant mass-to-light ratio is $M/L_K = 58\pm3 M_\odot/L_\odot$. The Virgocentric turnaround radius is $7.3\pm0.3$ Mpc. We explain several interesting trends in peculiar motions for various regions now that we can construct the 3D orbital histories.
\end{abstract}


\section{Introduction}

The large scale overdensity region of the universe that we reside in  was called the
Virgo Supergalaxy and later the Local Supercluster by \citet{1953AJ.....58...30D, 1956VA......2.1584D}. 
We focus on a domain of $\sim3000$ \kms\ $\sim 40$~Mpc. 
This article, on the dynamics of the region, is complementary to an article on
the groups and larger scale associations found in the
same region \citep{2017ApJ...843...16K}. 
The first extensive discussions of Local Supercluster dynamics date back 36 years \citep{Tonry1981, Aaronson1982}.
The Virgo Cluster is the main condensation in this region and its collective mass causes substantial peculiar motions.
The Centaurus and Hydra clusters, just beyond our sample
volume, and several other rich clusters collectively make up the ``Great Attractor" that is a significant source of a large scale flow enveloping the study region \citep{1987ApJ...313L..37D, Lynden-Bell1988}; see \citet{2017NatAs...1E..36H} for an accounting of attractors and repellers that affect local bulk motion. 
It has become evident that the so-called Local Supercluster is only an appendage of the Great Attractor complex which, in the fullness of its basin of attraction
is called the Laniakea Supercluster \citep{Tully2014a}.
Nearby components that call for our attention include
the Fornax Cluster and adjacent Eridanus Cloud that
\citet{1956VA......2.1584D} referred to as the Southern Supergalaxy, and the Local Void \citep{1987ang..book.....T}
that is actually part of an extensive network of voids.

Our interest is in galaxy flow patterns, departures from cosmic expansion. 
The radial component of these motions are decoded from
galaxy distance measures and the degree to which they deviate from the mean
Hubble law expectations. 

Distance estimates have become plentiful with the release of the \textit{Cosmicflows-3}
compendium of 18,000 galaxies \citep{2016AJ....152...50T}. The
almost 400 nearby, high precision, measures based on the tip of the red
giant branch method \citep{Jacobs2009} are particularly important for the present discussion.
We are able to recover the mass distribution from the observed motions of galaxies even though orbits in high density regions can be complex.  
Orbits manifest curvature and, across much of
the local volume, dynamics are in the non-linear regime, i.e. peculiar velocities are not linearly related to the overdensities.

At still higher densities, there are shell-crossings \citep{2012PhRvD..85h3005S}, however such situations are too complicated to be disentangled on the scales of current interest. Therefore, this work considers collapsed knots as indivisible units and refers to them as ``masses."

Before discussing our methodology, we mention other approaches to the mapping of mass in large-scale regions.  
Weak gravitational lensing (\citep{1992ApJ...388..272K, 1993ApJ...404..441K} provide mass projected onto the plane of the sky, and studies of cluster caustics \citep{Diaferio1997, Rines2003} provide total masses within some rich cluster.  

On large scales, where the fractional overdensities are low,
\begin{equation}
\delta \equiv \frac{\rho - \bar{\rho}}{\bar{\rho}} << 1
\label{eq:delta}
\end{equation} 
a simple linear relation holds between the divergence of peculiar velocities, $\boldsymbol{v_p}$, and the distribution of matter \citep{ 1990ApJ...364..349D}
\begin{equation}
\boldsymbol{\nabla} \cdot \boldsymbol{v_p}  = - H_0 f \delta
\label{eq:linear}
\end{equation}
where $\bar{\rho}$ is the mean global density, and the dimensionless velocity factor $f$ is defined in terms of the density growth factor $D$, approximated by \cite{Lahav1991} as:

\begin{equation}
f = \frac{a}{\dot{a}} \frac{\dot{D}}{D} \approx \Omega^{0.6}_m+ \frac{\Omega_{\Lambda}}{70}(1+ \frac1{2} \Omega_m).
\end{equation}

However, within a supercluster, fractional overdensities are not typically well below unity.
One option, used in the past for the nonlinear regime near clusters, has been to assume spherical infall \citep{Hoffman1980, Tully1984, Ekholm1999, Karachetsev_etal2014} in which each shell of matter follows the Friedmann equations of a standard universe of the same age, density, and $\Omega_{\Lambda}$.  
The actual situation is generally more complex than spherical infall, therefore realistic models need to be more elaborate.  \citet{Peebles1989} demonstrated a way forward for the more general case with the approach initially called `least action' but now called the `numerical action method' (NAM)\ \citep{Peebles2001, Phelps2006, Shaya2013}.  
The solutions are paths at either extrema or saddle points of the action, the integral of the Lagrangian over time, as test paths are varied arbitrarily.
NAM works with the comoving Lagrangian for collisionless particles, ie, their kinematic minus potential energies transformed into comoving coordinates.  

As in the usual celestial mechanics problem, six phase-space constraints are needed per particle to solve for particle motions influenced by their mutual gravity.
Since, at early times, peculiar velocities were small ($v_p \propto a^{3/2}$), consistent with a smooth 
homogeneous universe seen in the cosmic microwave background, NAM presumes for the first time step that peculiar velocities are consistent with linear theory.  
Since we now have fairly accurate distances to many galaxies, the other three constraints
can be fulfilled by the present 3D positions.

In NAM, each path is the path of the center of mass of the atoms of each final assemblage which, according to the center of mass theorem, moves independently of the internal motions, including inelastic collisions such as mergers.  
Hence, knowledge of the detailed history of assembly of each galaxy or group is, to first order, not needed.  
By assuming point or spherical particles, we
loose high order terms in the gravity field associated
with non-spherical shapes. 
These are only important during very close approaches or at very early times when overdensities were adjacent.
Our paths begin at redshift z = 4, before the present constituents were fully formed but after significant mass collapse had taken place.  

It is now evident that a significant component of peculiar velocities can be generated by large scale structure out to great distances.  
Nearby, the repulsion of the Local Void causes flow perturbations that are as important as Virgo attraction \citep{Tully2008, 2015ApJ...805..144K}.  
While Virgo infall might be our primary focus, a holistic approach is needed to attain a realistic history of any part of the Local Supercluster.  
As a first step, we construct paths over a region of 38 Mpc in radius centered on ourselves, including what we anticipate to be the full region of infall around the Virgo Cluster and all pertinent nearby attractors.  This region roughly coincides with what has historically been called the Local Supercluster. 
 
The Local Void extends to $\sim 5,000$~\kms\ and only if our model includes the {\it absence} of matter on this scale will the expansion of the Local Void boundary have the correct magnitude.  
In addition, the alignment of major masses, from the Shapley Concentration \citep{Raychaudhury1989, Scaramella1989} through the Norma$-$Centaurus$-$Hydra complex (Great Attractor)  \citep{Lynden-Bell1988}, to the Perseus$-$Pisces filament \citep{Haynes1988} creates a strong tidal field at our location \citep{1986ApJ...307...91L, 1999ApJ...522....1D, 2005A&A...440..425R}.  
The large scale density field has recently been studied by \citet{Tully2014a} and the three-dimensional map of structure generated by that study is used here to derive a prescription for these external influences.

It is important to appreciate what a modeling of galaxy paths can and cannot accomplish.  The paths that are recovered are physically plausible, but they are not unique.  
The search for a path for any particle can lead to multiple solutions, particularly in regions that collapsed at early times.  
As a simplification, the particles that we consider in the orbit constructions may contain multiple galaxies, which allows us to avoid calculating paths within the virial inner region of groups and clusters.
Even so, usually there are multiple solutions per mass particle. 
We try to select only simple paths; if a path is complicated, or if it is wildly off in $v_m$, then we can quickly search to see if a better path is available by a technique we call backtracking.  We solve for paths satisfying the velocity constraints at the earliest time step by iterating backwards in time from the present position and $v_o$, repeatedly to cover in a tight grid, all directions for the present velocity.  If we find another path that is uncomplicated in this process, then we switch that one in and rerun the NAM process to allow all paths to resettle.

It is unlikely that we find the correct path for every mass, but presumably, 
if the model fits well, a majority of the orbits are qualitatively correct and the input parameters are close to correct.  Tests of NAM using N-body simulations on a range of scales confirm these basic tenants \citep{1994ApJ...434...37B, 2000MNRAS.313..587N, 2002MNRAS.335...53B, Phelps2006}.

Here are some questions that we posed that this modeling could answer.    
What are the total masses to be associated to halos on the scales reaching all the way until the neighboring halo's mass distribution begins?  
Cluster masses have been measured by their velocity dispersions with the virial theorem, but how much mass associated with a cluster is present beyond the volume measured by the virial theorem?  
Where is the Virgo Cluster turnaround radius today?
Where is the gravitationally bounded limit, and what fraction of the total supercluster volume is associated with these domains?
What fraction of matter is presently unassociated with galaxies, either still in the primordial intergalactic medium or tossed out by gravitational encounters, ram pressure or internal explosive activity? 
Then, what are the dominant flow patterns besides Virgo infall?  
Streaming toward the Great Attractor and away from the Local Void have been mentioned.  
These motions are the most obvious but there are other coherent patterns that deserve attention.

In \S 2, we discuss some of the new observational data that has accumulated recently that benefit this study.  
In \S 3 we discuss the data and analysis that went into the determination of the external field arising from 
inhomogeneity beyond the volume of the orbit study.  
A description of the flow model and the $\chi^2$ 
goodness-of-fit criterion based on model and observed velocities is given in \S 4. 
The best model after $\chi^2$ minimization by varying input parameters is discussed in \S 5.  

The Discussion section, \S 6, presents a summary of results and implications.

\section{Observational Input} \label{observations}
 
We separate the analysis into two domains: an inner spherical core where we follow orbital paths and an outer domain of comoving, growing tidal influences. 
Within the core volume, centered at our position, paths are reconstructed for 1,382 individual galaxies and galaxy groups. 
The dominant individual constituent within the core is the Virgo Cluster at a distance of 16 Mpc.  An important secondary feature is the cluster called Virgo W  \citep{deVaucouleurs1961} only slightly off the line of sight behind the Virgo Cluster at 33~Mpc.  The choice of a transition between the inner and outer domains of 38 Mpc, $\sim 2,850$~\kms, comfortably includes the Virgo W Cluster and lies at a radius where there is little structure (the exception is in the direction of the Centaurus Cluster that lives just outside the core region).

With NAM we can find the paths that were taken for all the dynamically important masses plus those masses, 
even light ones, with accurately known distances.  
All massive halos must be followed for a realistic model, whether or not an individual case has a 
good quality distance estimate. 
Therefore, for masses lacking quality distances, we adjust distances as needed until the model 
redshift agrees with observations.  These will not be included in the goodness measures. 
All masses with well determined distances, even if they have an inconsequential mass, are useful as test particles that sample the gravitational field and can contribute to the goodness criteria.   
Consequently, our model contains a volume complete description of the visible mass, dominated by the more massive components and optimized to produce accurate velocities for the masses with quality distances.

In addition to the phase-space constraints, the numerical action path reconstructions require a proxy for masses - here provided by 2MASS $K_s$ luminosity  \citep{Jarrett2000} multiplied by a mass-to-light ratio parameter to be optimized.  
The 2MASS redshift survey \citep{Huchra2012} provides $K_s$ magnitudes and is essentially complete in redshifts for galaxies 
brighter than 11.75 mag in that band over the entire sky except within $5^{\circ}$ of the 
Galactic plane.  
The characteristic limit of the 2MRS11.75 catalog (peak in the number counts) is $\sim 10,000$~\kms, 
well beyond the depth of our core region.  
Galaxies in the 2MRS catalog have been assigned to groups  \citep{Tully2015} giving collective 
masses that are of interest for the present discussion. 
With increasing distance, we group more drastically by aggregating nearby associations into 
single items. After this grouping procedure, individual masses beyond distance modulus 
$\mu$ = 29.22 mag and with luminosity $< 10^{9.5} L_\sun$ or beyond $\mu$ = 32.6 and with luminosity $< 10^{10.3} L_\sun$ were dropped to reduce computation time.  
As mentioned, the 2MRS11.75 catalog excluded the zone within 5\degr\ of the Galactic plane. 
This zone of avoidance was filled in by drawing on the main 2MASS photometry catalog \citep{Jarrett2000}.  
About 100 relevant galaxies were found with $|b| < 5\degr$ and these were combined into 40 groups.   Between observations in the infrared and blind HI surveys \citep{1987ApJ...320L..99K, 2016AJ....151...52S} there can be reasonable confidence that no major nearby dynamical influence has escaped detection.

The other ingredient needed is good distances.
The accepted distance to a group is the weighted average of the available distance moduli.  
The radial velocity component in the rest frame of the center of the MW, $v_o$, of the group is taken to be the unweighted average of all members (whether or not a distance is known).  
The individual distances are reported in the \textit{Cosmicflows-3} data release \citep{2016AJ....152...50T}.  
They are derived from six alternative methodologies of differing accuracies.  Highest accuracies ($\sim 7\%$) are achieved using either the period$-$luminosity relation in the pulsation of Cepheid stars \citep{2016ApJ...826...56R}, the constancy of the luminosity of the tip of the red giant branch \citep{Lee1993a}, or (in a small number of cases) the standard candle nature of supernovae of type I$a$. 
Currently almost 400 high precision distance measures are available, giving dense coverage of the local region with high accuracy.  
The surface brightness fluctuation method \citep{Tonry2001} provides a particularly valuable component for targets observed from {\it Hubble Space Telescope} in the Virgo and Fornax directions  \citep{Blakeslee2009}, with uncertainties $\sim 10\%$.  Then, there are plentiful, but individually less accurate measures, coming from the correlation between spiral galaxy luminosity and rotation rate \citep{Tully1977} and the related fundamental plane correlation between luminosity, central velocity dispersion and surface brightness or size for elliptical galaxies  \citep{1987ApJ...313...42D, 1987ApJ...313...59D}, each uncertain at the level of $\sim 20\%$ (See the 
\textit{Cosmicflows-3} reference for details).  

\begin{figure}
\plotone{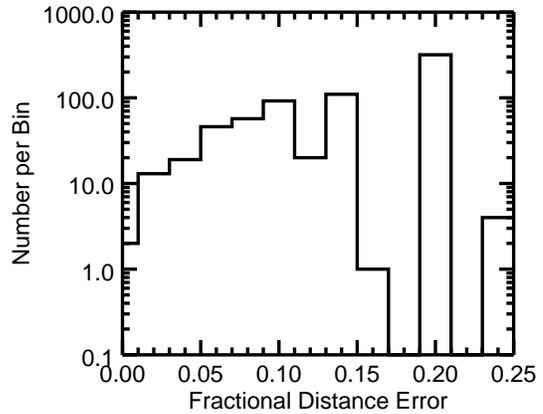}
\caption{\textbf{Histogram of the number of masses in  bins  of fractional distance error} after weighted averaging of member distance measurements. Only masses with distance uncertainties below 15\% were used within $28$ Mpc and only uncertainteis below $10\%$ beyond that for the $\chi^2$ statistic.\label{fig:derr}}
\end{figure}

Averaging of distances among the members of groups diminishes uncertainties, sometimes dramatically.  
The nominal group uncertainties from the summed weights $-$ only 1\% for Virgo and as low as a few percent in other good case $-$ are surely overtaken by systematics.  
After grouping, the fractional distance errors of the sample have the distribution shown in the histogram of Fig.~\ref{fig:derr}.  
A total of 681 masses have distances, but we constrain the model fit using only the 264 with distance errors $\le 15\%$ and within 28 Mpc or $\le 10\%$ at $28-38$ Mpc.  
The 1,382 masses tracked in the numerical action analysis are listed in Table~\ref{tbl_rslts}. 

\section{External Field}
\label{sec:external}

The gravitational effects induced by inhomogeneities beyond the sampled region make a substantial contribution to the total gravitational field that generates peculiar velocities. 
Evacuation from the Local Void plays an important role in determining both the amplitude and direction of peculiar velocities \citep{Tully2008c}.  Although irregular in shape and not entirely empty, the Local Void is as close as the edge of the Local Group and extends to $\sim 5,000$~\kms$\sim 70$~Mpc from us.   
In addition there are such distant overdensities as the Great Attractor tugging on us.  
The influence from density fluctuations beyond the core is represented by a gravity field that changes over time only in amplitude in the comoving frame.  

For the external field, we rely on studies based on the \textit{Cosmicflows-2} survey of $\sim$8,000 galaxy distances and velocities with $z < 0.1$ \citep{Tully2013}. 
The survey included mostly luminosity-linewidth relation distances for spirals and fundamental plane distances for E/S0 galaxies, with contributions on large scales from $\sim 300$ supernova distance measures.   
The data were analyzed to recover both the 3D peculiar velocity field and the overdensity distribution using the Wiener Filter method \citep{Zaroubi1995a, Zaroubi1999, Courtois2012}.  
The Wiener Filter recovers the components of flows generated by regions extending to $z \sim 0.05$, providing a description of influences relative to the center of mass generated on all relevant spatial scales.  

The Wiener Filter reconstruction of \textit{Cosmicflows-2} was used to create the external field in three steps.
First, the peculiar velocity and density fluctuation fields were generated by means of the Wiener Filter, producing 3D maps of $\boldsymbol{v}_{WF}$ and $\delta_{WF}$.  Second, 
the velocity field due to the mass within $R=38$ Mpc,  was then calculated by solving the Poisson equation for the density field for $r<R$.
Third, the difference was taken between the Wiener Filter and local fields to construct the external field: $\boldsymbol{u}_{ext} = \boldsymbol{u}_{WF} - \boldsymbol{u}_{local}$.
The map indicates an externally generated flow at the Local Group position of 255 \kms\ in the cosmic microwave background frame, partitioned in the cardinal supergalactic directions $v_{SGX},v_{SGY},v_{SGZ} =[-212, 95, -106]$ \kms. 
The center of mass velocity is removed from the entire map since, according to the equivalence principle, a uniform acceleration is equivalent, internally, to a space free of gravitational fields and hence has no effect on paths in the sample's reference frame.
  
The resulting $v_{ext}$ map, sampled at 1 Mpc intervals in 3 dimensions was fit by a third order polynomial in three dimensions to allow for interpolation at any location and to make it possible to form a first derivative function of the gravity needed by NAM.  
It is presumed that $\boldsymbol{u_{ext}}$ developed from the external gravity field and grew according to linear perturbation theory.
The time dependence of the comoving  acceleration, $\boldsymbol{g_{ext}}$, arising from sources external to 38 Mpc can be surmised from the following \citep{1980lssu.book.....P}
\begin{equation}
\boldsymbol{g_{ext}}(\boldsymbol{x}) \equiv  -\frac{\boldsymbol{\nabla}\phi}{a} = G a \int_{ext} d^3\boldsymbol{x}^\prime \delta  \bar{\rho} \frac{\boldsymbol{x}^\prime - \boldsymbol{x}}{| \boldsymbol{x}^\prime - \boldsymbol{x}|^3}
\end{equation}
where $\phi$ is the gravity potential, $a$ is the global expansion factor, $\boldsymbol{x}$ is a comoving position, and the integral is taken over all space outside of the core region.  
Since $ \delta \sim a$ and $\bar{\rho} \sim a^{-3}$, then $g_{ext} \sim 1/a$.  
Using the relation for present peculiar velocities given present acceleration \citep[p.~64]{1980lssu.book.....P}
\begin{equation}
\boldsymbol{u}_p = \frac{2}{3}\frac{f \boldsymbol{g}_0}{H_0 \Omega_m}
\end{equation}
where $f$ is given in Eq. 2,
the external comoving acceleration can be expressed as 
\begin{equation}\label{eq:gext}
\boldsymbol{g}_{ext}(t) = \boldsymbol{g}_0/a(t) = \frac{3 H_0 \Omega_m}{2a(t)f}\boldsymbol{u}_{ext}. 
\end{equation} 

\section{Flow Model Description} 

The NAM technique comes in two flavors; one that assumes known distances, solves for paths and outputs model redshifts, and another that assumes known redshifts and solves for the paths, including present positions \citep{Phelps2002}.  
Since many distances are now accurately known in the region of study, while along some lines-of-sight there are multiple locations with the same redshift, we exclusively used the procedure that assumes distances. 
In particular we use the NNAM version of NAM described in detail in the appendix of  \cite{2010arXiv1009.0496P}.
In the cases where distances are not known, we simply allow the distances to drift, after settling on a reasonable set of input parameters, until the output redshifts agree with the observed one.  
All paths have 100 time steps with uniform spacing in $a$, from z=4 to z=0.
A flat universe, ie, $\Omega_\Lambda=1-\Omega_m$ is assumed in all models.  

To aid in obtaining the simplest set of paths, at the outset we made a set of runs that began with an extremely low mass-to-light ratio and gradually increased the value, taking the results of each run for the input set of paths of the next run.  
Meanwhile, the value of $\Omega_\mathrm{orphan}$(defined in the next paragraph) began the runs nearly equal to $\Omega_m$ and gradually decreased to compensate for the increase in galaxy masses.  
This procedure, allowed the paths to gradually grow up from nearly zero length and  successfully resulted in few tangled paths.
 
The following parameters are adjusted to arrive at a minimum for the goodness criterion:  the masses of the Virgo Cluster, the MW, and M31; either a single mass-to-$Ks$ band light assumption for all other masses or a slightly more complex formulation to be discussed;  and a constant density component that crudely represents all dispersed particles $\Omega_\mathrm{orphan}$.  

In addition, the amplitude of the circular velocity of the local standard of rest around the MW, $\Theta_\sun$, is an important parameter because it is reflected in all other masses. 
We nominally chose 239 \kms\ \citep{Marel_etal2012}.
The motion of the Sun in the the Local Standard of Rest is small and well enough determined; we set it to $V_\sun^{LSR}$ = (9.8, 11.6, 5.9) \kms\ \citep{Jaschek1992} in galactic u,v,w coordinates, or (0.6, 4.2, 15.7) \kms\ in supergalactic Cartesian coordinates.
 
\subsection{Goodness Criterion}
The peculiar velocities ($\boldsymbol{u} = a\dot{\boldsymbol{x}}$) at the end of the NAM trajectories are projected onto the line of sight to form the present model redshifts as seen from the center of the Galaxy,
\begin{equation}
v_m = (\boldsymbol{u} - \boldsymbol{u_{MW}}) \cdot \boldsymbol{\hat{x}} + H_0 d.
\end{equation}
The model redshifts are compared to observed redshifts, $v_o$, in a $\chi^2$ goodness statistic using
only masses with distance errors $<15\%$. 
For groups, the uncertainty in $v_o$ is taken to be the standard deviation of the mean of the members' velocities, $\sigma_{e}= \sigma_v/\sqrt{N}$ while for an individual galaxy an error of 20 \kms\ is assigned.  
The goodness criterion is a $\chi^2$ of the deviation between model $v_m$ and observed $v_o$ weighted by the expected error from the combined distance errors and velocity errors.  
An additional error of $\sigma_{dm}$ = 20~\kms\ is added in quadrature to account for the possibility that the velocity of baryons of a galaxy, group or associations may depart from the mean velocity of the dark matter distribution.  Hence

\begin{equation}
 \chi^2 =\frac{(v_o-v_m)^2}{(H_0 e_d d)^2+\sigma_{e}^2+ \sigma_{dm}^2}
 \label{eq:chi}
\end{equation}
where $e_d$ is the fractional error in distance.

A simple minimization scheme is used in which each parameter is adjusted by taking fixed sized steps until a minimum in $\chi^2$ is reached and then the next parameter is adjusted.

After each loop, the step sizes are reduced by half until going through a loop reduces $\chi^2$ by an insignificant amount.
Gradual adjustments are preferred for this problem because paths have multiple solutions,  and jumps to alternative solutions, during action minimization, are more likely when parameters are changed too rapidly.  
For each step in $\chi^2$ minimization, the run is begun with the same initial paths to prevent comparing cases with alternate path choices. 

Unfortunately the automated procedure does not always traverse parameter space to the absolute minimum on its own.  
Spontaneous path jumping can occur at any time and this often leads to a cascade in which many other masses jump to different paths, sending the $\chi^2$ either up or down by several sigma.   
Because $\chi^2$ is discontinuous in the input parameters, the minima may be only local minima. 
Therefore the procedure needs to be restarted with many different parameter settings.  
Although we tried nearly 100 different starting points, we cannot guarantee that the search was completely exhaustive and that the absolute best set of values has been found.

\section{Best Model}

We assumed \Ho\ = 75 \kmsMpc\ for the majority of our analysis to be consistent with the ensemble of distances used in creating our input catalog,  {\it Cosmicflow-3}  \citep{2016AJ....152...50T}, appreciating that a shift in the zero point of the distance scale set locally does not affect a peculiar velocity analysis.  The discordance with the value $H_0 = 68$~\kmsMpc\ determined from the Planck analysis is noted  \citep{2016A&A...594A..13P}.

\begin{deluxetable}{cccccccc}
\tablecolumns{8} 
\tabcolsep=0.11cm
\tablecaption{Best Models  \label{tbl:models}} 
\tablehead{ 
\dcolhead{H_0} & \dcolhead{\chi^2} & \dcolhead{M_0}     
& \colhead{M/L}           & \colhead{Virgo}       & \dcolhead{\Omega_{orp}} & \dcolhead{M_{MW}}    &  \dcolhead{M_{M31}}\\
\nocolhead{one}  & \nocolhead{chi2} & \nocolhead{}  
& \dcolhead{\frac{M_\sun}{L_\sun}} & \dcolhead{\frac{M_\sun}{L_\sun}} &\nocolhead{omegao} & \dcolhead{10^{12} M_\sun}  & \dcolhead{10^{12} M_\sun}
}
\startdata
\cutinhead{Mass $\propto$ Light$^{1.15}$}
75 & 0.393 & 39.6 && 113 &  0.077 & 2.29 & 2.86\\
70 & 0.419 & 33.7 &&  97 & 0.040 & 2.38 & 3.06 \\ 
67 & 0.533 & 33.1 && 105 & 0.000 & 2.46 & 2.56\\ 
\cutinhead{Mass $\propto$ Light}
75 & 0.400 && 57.8 & 113 & 0.078 & 2.37 & 2.86\\ 
70 & 0.462 & & 49.9 & 127 & 0.053 &2.42&3.32\\
67 & 0.515 &&  47.9 & 109 &  0.000 & 1.86 & 2.84\\ 
\cutinhead{ $1 \sigma$ error estimates}
partial & 0.008& 0.36 & 0.23 & 2.5 & 0.0012  &0.35 &0.35  \\
full & 0.008  &   2.04 & 3.00 & 15.1&  0.016 &0.75 & 0.75\\
\enddata
\end{deluxetable}

The mass-to-light assignments are a key parameter and we experimented with two variants: a constant with luminosity and a power law with luminosity. 
We are guided by literature evidence \citep{Marinoni2002, Tully2005, 2009ApJ...695..900Y}
of a mild increase in the mass-to-light
ratio with increasing halo mass with exponent 0.15 from virial mass measurements. We consider the relation

\begin{equation}
M/L_{K_s} =   M_{0} ~ L_{10}^{0.15}  ~M_\sun/L_\sun,
\label{eq:mtol}
\end{equation}
where $M_0$ is a normalization constant and $L_{10}$ is the $K_s$-band luminosity in units of $10^{10} L_\sun$. 
This shallow power law formulation led to a slight improvement in goodness of fit. 
We will mostly present results and figures with our solution with this mild power law growth in mass-to-light.

\begin{figure*}
\plotone{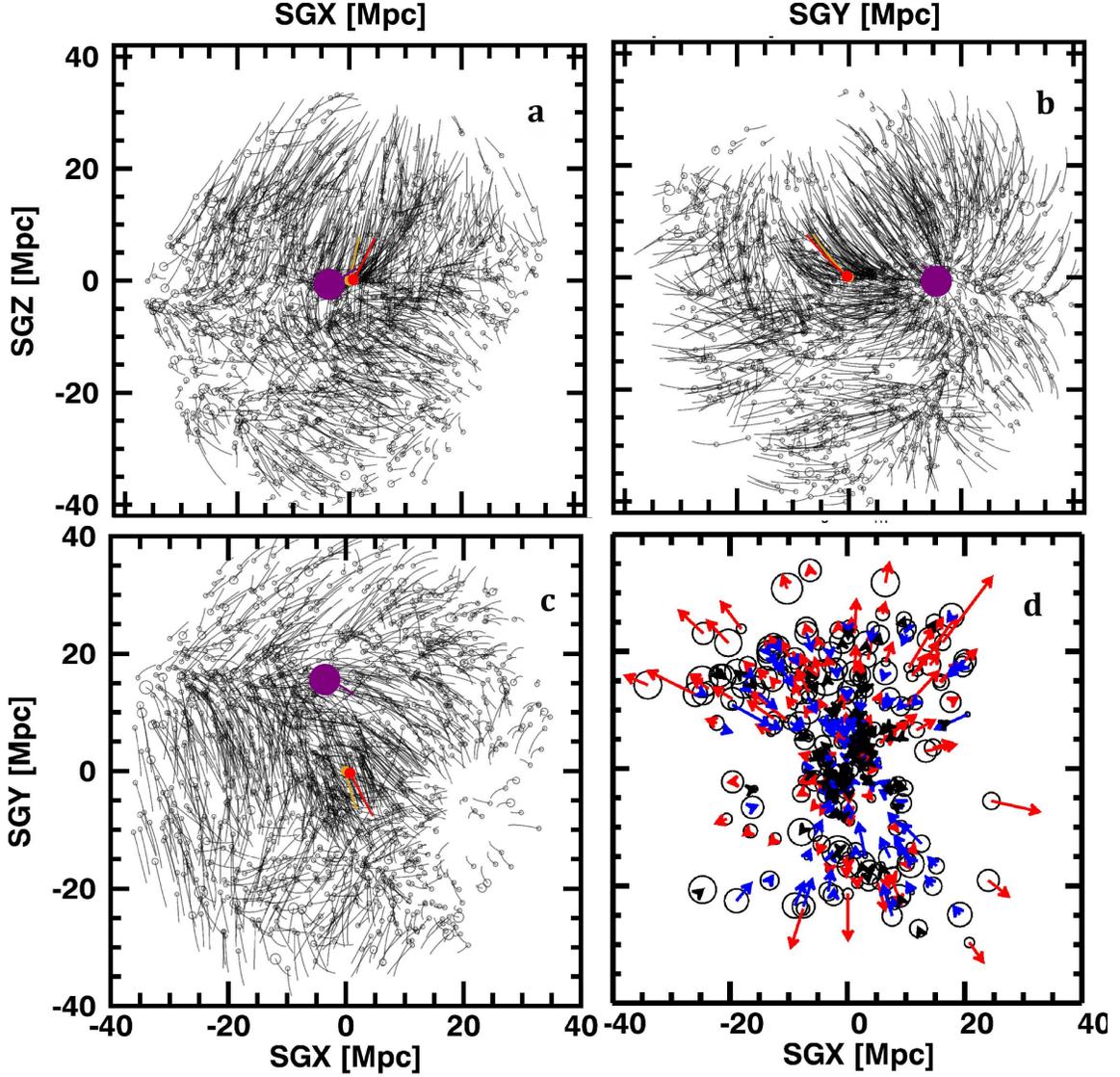}
\caption{\textbf{Paths of projected onto supergalactic cardinal planes} from z=4 to the present in comoving coordinates.  \textbf{a}: SGX-SGZ, \textbf{b}: SGY-SGZ, \textbf{c}: SGX- SGZ. Circles are placed at present positions sized in proportion to the logarithm of the ascribed mass. MW is gold, M31 is red, Virgo Cluster is purple. The reference frame is the center of mass of core sample. \textbf{d: Observed minus model redshift map} shows projection in XY plane of radial vectors of $v_o - v_m$ for objects with quality distances used in $\chi^2$ evaluation.  Arrows are blue: $v_o - v_m < -35$, black: $-35 < v_o-v_m < 35$, and red: $v_o-v_m > 35~ $\kms.} 
\label{fig:orbitsxyzcz}
\end{figure*}

The models show the Local Group strongly flowing both out of the Local Void and towards the Virgo Cluster on fairly simple paths, as do most of the other galaxies in the local plane of galaxies (Fig.~\ref{fig:orbitsxyzcz}a-c). 
The motion away from the Local Void shows no signs of abatement.

Table~\ref{tbl:models} presents results after minimizing the median in
$\chi^2$ with various choices of \Ho\ (discussed in \S5.1).  
It provides the minimized $\chi^2$ value, 
either the normalization constant for a power law in mass-to-light ratio or a best constant value, a mass-to-light ratio for the Virgo Cluster,
$\Omega_\mathrm{orphan}$, and masses for M31 and the MW.  
At the bottom of the table are two rows with 1 standard deviation estimates, $\sigma$, for both 'partial' and 'full' variations, 
where partial means that only the single parameter is varied and 'full' is where all other parameters are allowed to vary to bring $\chi^2$ down again.
Nominally $1 \sigma$, with 286 free parameters, would be equivalent to change of 0.02 in $\chi^2$.  
However, the median has an expectation of 0.997 and a 95\% chance of being over 0.865.  
We interpret the low median $\chi^2$ to the non-Gaussian nature of the errors in the distance relation, which resulted in error assignments too high when working with only the best half of the distribution.  
To compensate for this, we have reduced the $1 \sigma$ error assignment to 0.008 or $\sim 0.02\chi^2$.  

The best model with \Ho\ = 75 \kmsMpc, $\Omega_m = 0.137/h^2 = 0.244$ ($t_0=13.30$ Gyr), and an exponent of 0.15 in the mass-to-light relation of Eq.~\ref{eq:mtol},
occurs with the following parameters:  $\Omega_\mathrm{orphan} = 0.077\pm0.016$, $M/L_{K_s} = 40 \pm 2~L_{10}^{0.15}~M_\sun/L_\sun$ ($L_{10}$ is K-band luminosity in units of $10^{10}~L_\sun$), a Virgo mass of $6.3\pm0.8 \times 10^{14}~M_\sun$ ($M/L_K = 113\pm15~M_\sun/L_\sun$), and a sum for the mass of M31 and MW of $5.15\pm0.35 \times 10^{12}~M_\sun$.  The latter is in close agreement with a recent result by \cite{2016arXiv160602694M} of $M(Local Group) = 4.9\pm0.8 \times 10^{14}~M_\sun$.  There are large 'full' error bars on the M31 and MW individual masses, primarily caused by the weak constraints on how this mass was split between the two galaxies.  
Individually, the model masses for the MW and M31 are 2.29 and $2.86\pm0.75 \times 10^{12} M_\sun$, respectively.

The coefficient on $L_{10}^{0.15}$ in the best model is $M_0=40$, which is close to the value of 32 found by \citet{2015AJ....149...54T}.\footnote{The coefficient of 43 given in \citet{2015AJ....149...54T} assumed $H_0 = 100$~\kmsMpc.}  The best run with constant mass-to-light ratio was worse but only by about $1 \sigma$ and had $M/L_{K_s} = 58 \pm 3~M_\sun/L_\sun$.

Fig.~\ref{fig:orbitsxyzcz}d is a vector diagram of the residual of redshifts after subtracting the model redshifts  projected onto the SGX-SGY plane for the 286 masses with precision distance determinations.  The arrows have length of $\frac{1}{H_0}(v_o - v_m)\hat{r} \times \hat{z}$ and are colored black if $v_o$ and $v_m$ agree within 35 \kms, red  if $v_o > v_m + 35$, and blue if $v_o < v_m - 35$ .  The high number with good agreement and the lack of any strong regional trends (after excluding a few zingers), strengthens the case that the model is not missing any crucial components and gravity was indeed the origin of the peculiar motions and  structure.

\begin{figure}
\epsscale{1.2}
\plotone{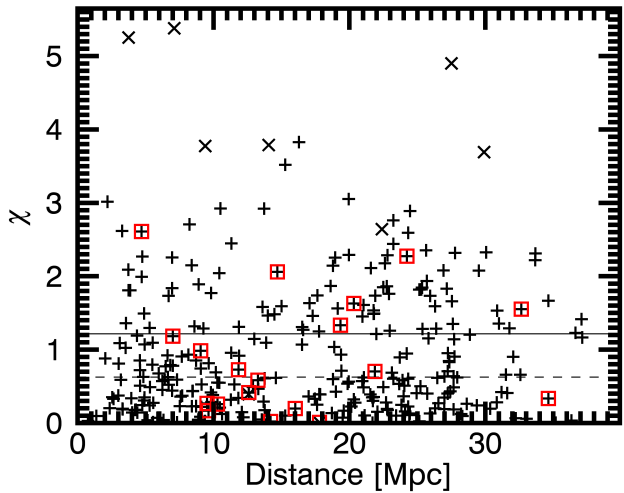}
\caption{\textbf{Individual $\chi_i = (\chi_i^2)^{1/2}$ values} in the best model, with power law mass-to-light ratio, for masses with high quality distances. Crosses were outliers in all models and therefore excluded in the final statistics. Pluses are included in the mean (solid line) and median (dashed line). A square is added to masses in the 15\degr\ cone about the Virgo Cluster}.
\label{fig:DvsChi} 
\end{figure}
\begin{figure}
\epsscale{1.15}
\plotone{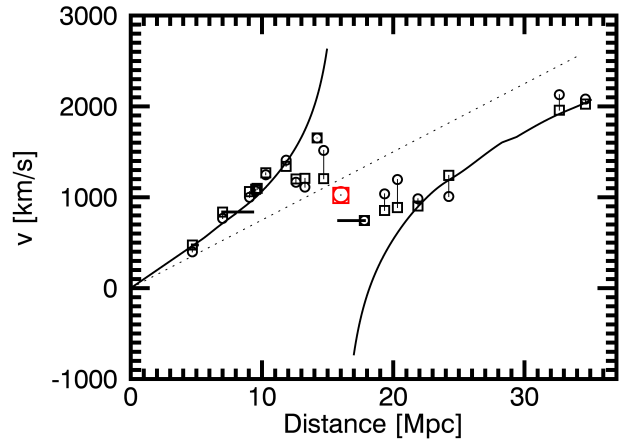}
\caption{\textbf{Distance vs velocity for masses in the 15\degr\ cone} centered on the Virgo Cluster.  Model $v_m$ (squares) are connected to the observed $v_o$ (circles) by vertical lines.  
If a distance was   adjusted to improve the fit, the displacement is shown as a horizontal line. Dotted line is $H_0=75$ relation. Thick solid line shows $v_m$ for test particles along line to the Virgo Cluster.}\label{fig:DvsV15}
\end{figure}

In Fig.~\ref{fig:DvsChi}, the contribution from each object to the $\chi^2$ is presented  as a function of the distance from us.  
Masses with $\chi_i > 3$ in all models were added to a blacklist and excluded from the goodness calculations.
These masses could be in complicated orbits or the data could be erroneous. 
The red squares in Fig.~\ref{fig:DvsChi} highlight masses that lie within the 15\degr\ cone of the Virgo Cluster, and just these are plotted in 
Fig.~\ref{fig:DvsV15}, which shows $v_o$ and $v_m$ versus distance.

We were surprised that there is no evidence for significant additional mass beyond the Virgo Cluster's virial mass. 
Our mass is intermediate between the virial mass determined when using only early 
type galaxies of $4 - 5 \times 10^{14} M_\sun$ and the mass determined using all members of 
$\sim 9 \times 10^{14} M_\sun$ \citep{Tully1984}.  
It would seem that the spirals inside the cluster are mostly still on radial orbits rather than virialized into an isotropic velocity distribution.
The mass-to-light ratio of the cluster, $113 \pm 15 M_\sun/L_\sun$, is quite close to the value of 103 predicted by our relationship $M/L_{K_s} = 40 \pm 2~L_{10}^{0.15}~M_\sun/L_\sun$.

In this model $v_m$(M31) = $-106$~\kms\ is in very good agreement with the observed value 
of $-107$~\kms\ ($\Theta_\sun$= 239 \kms).  
The very low proper motion of M31, $4~\mu\rm{as~yr}^{-1}$ is compatible with proper motion measurements 
\citep{2012ApJ...753....7S, Marel_etal2012}.  
This result required a choice between three families of solutions involving the MW and M31: M31 falling rapidly in SGZ while the MW moves mostly along the supergalactic plane, MW falling rapidly in SGZ while M31 move mostly along the plane, and both MW and M31 falling moderately in SGZ.  
We selected only the last family to study, which always had a low proper motion for M31 and thus lower masses, because the best $\chi^2$ values were found among these solutions.  

Table~\ref{tbl_rslts} presents results of velocities and masses in the best model.  
The first column gives the name of the most prominent member of the system of galaxies.
In the second column, the IDs that start with the digits 1 and
2 refer to nests (groups) of galaxies from \cite{2015AJ....149...54T},
while entities that are additional to that reference have
an ID beginning with 3. The velocities $v_o$ and $v_m$ are
observed and model radial velocities, respectively. 
Distances for masses with zero in the distance error lack precision distance estimates; in these cases distances 
were adjusted to bring model velocities into agreement with observed velocities. In cases with quoted errors, distances are fixed by their measurements.  
Distances unconstrained by measurements were adjusted once, when the input parameters became reasonable, and were not readjusted after best parameters were found to ensure a completely self consistent model. 
Thus, velocities are not in exact agreement even for masses that were moved to get agreement.

The uncertainties in the last column are fractional uncertainties in
the distance.
Here are the first 10 rows of our results table.  The full table can be found online.

\begin{figure}
\epsscale{1.2}
\plotone{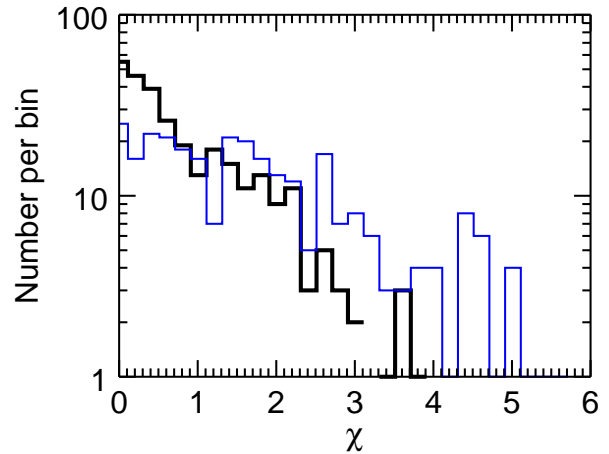}
\caption{Histograms of model goodness criterion $\chi^2$ (black and thick) and the same for Hubble velocities based just on distance (blue and thin)}\label{fig:chi_hist}
\end{figure}

\begin{deluxetable}{lrrrrrr}
\tabcolsep=0.11cm 
\tablecolumns{7} 
\tablewidth{0pc} 
\tablecaption{Results Table.  \label{tbl_rslts}} 
\tablehead{ 
\colhead{Name}    &  \colhead{ID} &   \colhead{Mass}   & 
\dcolhead{v_o} & \dcolhead{v_m}   & \colhead{d}    & \dcolhead{e_d/d}\\
\nocolhead{}    &  \nocolhead{} &   \dcolhead{10^{11} M_\sun}   & 
\dcolhead{km/s} & \dcolhead{km/s}   & \colhead{Mpc}    & \nocolhead{}
}
\startdata 
  MW&300000&  22.87&   &   &  & \\
M31&300001&  28.62& -106& -107&  0.8& 0.02\\
P3804975&103433&   0.00&   73&   61&  1.0& 0.00\\
N3109&112997&   0.02&  153&   28&  1.4& 0.00\\
U4879&300080&   0.00&   16&   19&  1.4& 0.10\\
N55&300202&   1.20&  133&  154&  2.0& 0.03\\
U9128&300028&   0.00&  122&  196&  2.2& 0.04\\
I3104&300206&   0.05&  226&  233&  2.4& 0.10\\
I4662&212362&   0.05&  187&  199&  2.5& 0.10\\
U8508&300024&   0.01&  175&  185&  2.6& 0.08\\
&&&&&& \\
...\\
\enddata 
\begin{verbatim}
Resulting model velocities for best NAM model.  
First 10 rows are shown here.  The entire table 
is available at: 
http://www.astro.umd.edu/~eshaya/action_flow/
result_table.pdf 
\end{verbatim}
\end{deluxetable} 
It is useful to quantify the improvement of using this peculiar velocity flow model over just assuming Hubble flow because we use the model to determine distances to masses with no distances. 
Fig.~\ref{fig:chi_hist} is a histogram of the distribution of $(\chi^2)^{1/2}$ values for the 286 precision distance masses (thick,black) and the same distribution but relying on  $H_0d$ for model velocities (thin,blue).  
The Hubble flow derived median of $\chi^2$, using Eq.~\ref{eq:chi}, is 2.65 versus the NAM-based value of 0.393.  

The orbits in our preferred model can be seen in the animated Fig.~\ref{fig:animation} and, alternatively, the associated interactive figure.  The Virgo, Fornax, Antlia and Virgo~W clusters are represented by large spheres colored red, gold, black, and purple, respectively.  The Milky Way and M31 are shown by smaller spheres colored yellow and green, respectively, in the animated figure and yellow and green, respectively, in the interactive figure.  The remaining of the 1382 masses considered in the orbit construction are colored blue and have sizes proportional to their masses.  There is an abundance of small spheres within 10 Mpc of the Milky Way.  These typically are modest galaxies with negligible gravitational impact but have accurate distances from tip of the red giant branch measurements.  They are important test probes of the potential field.  By contrast, some of the more massive elements represented by larger symbols do not have accurately determined distances but need to be included (at distances that accommodate their velocities) to properly represent the mass in halos in the study volume.

\begin{figure}
\epsscale{1.2}
\plotone{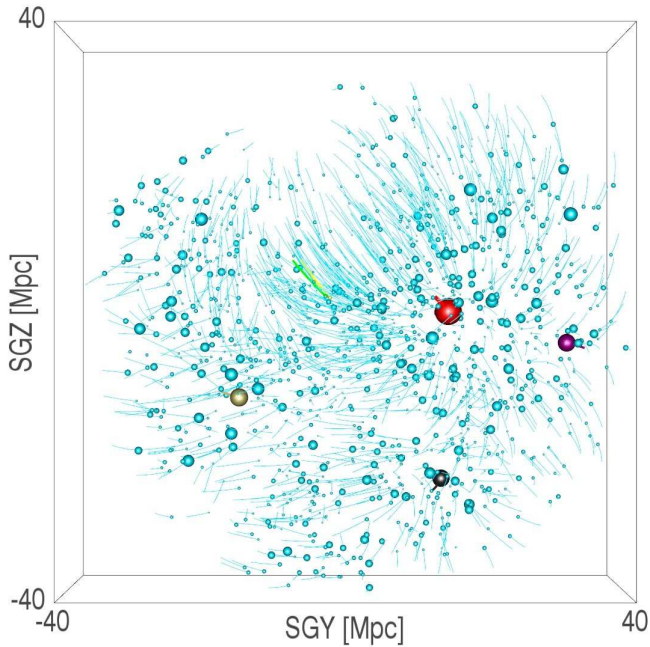}
\caption{\textbf{Online animation} of the orbits of masses within 38 Mpc.  A high resolution version of the film can be found at https://vimeo.com/239075970 
  .}
\label{fig:animation}
\end{figure}

\begin{figure}
\epsscale{1.2}
\plotone{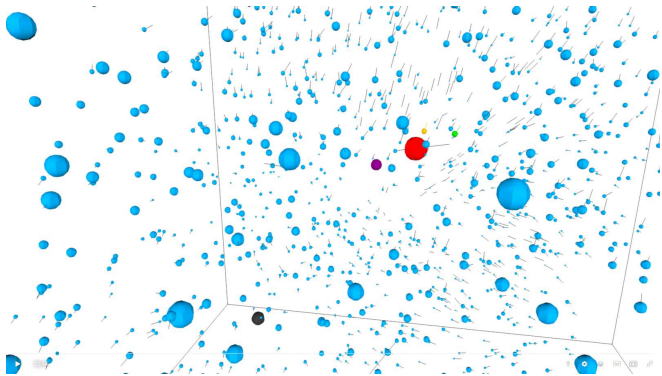}
\caption{\textbf{Online interactive 3D figure} of the orbits of masses within 38 Mpc can be found at https://skfb.ly/6tyxW .}
\label{fig:animation}
\end{figure}
There are a small number of masses with very divergent orbits to be seen in the animated figure.  Typically these are cases that have experienced a spurious close encounter at an early time in the construction and should be ignored.  Indeed, no individual orbit should be considered as better than plausible.  However the bulk features of our preferred model are shared by all variants with low $\chi^2$.  In particular, mass assignments are similar for the best models.

We draw attention to four-dimensional features of the interactive figure.  
Controls are not only spatial (free rotation, zoom, scale) but also temporal 
(play/pause, scroll forward/backward in time).

\subsection{Variation with \Ho}

In addition to \Ho\ = 75 and $\Omega_m = 0.244$, we also ran models with \Ho\ = 70 and $\Omega_m = 0.279$, satisfying the WMAP 5 year result \citep{2009ApJS..180..330K} that $\Omega_m h^2 = 0.137$, which fixes the present global density as \Ho\ varies.
Summary results are tabulated in Table~\ref{tbl:models}.
 As \Ho\ decreases, the age of the universe increases, but the increase in $\Omega_m$ moderates the age change to only 5.4\% .  

We can examine Newtonian gravity scaling relations \citep{Shaya1995} to get an idea of how one can change parameters to take paths that one knows are solutions and rescale them into new solutions, albeit not necessarily at the same $\chi^2$. If time, positions, and masses are rescaled via:
$$
t^\prime = \alpha t,\quad \boldsymbol{x^\prime} = \beta \boldsymbol{x},\quad m^\prime = \gamma m
$$
then rescaled paths are solutions as long as $\gamma = \beta^3/\alpha^2$ is satisfied, and velocities scale as $\boldsymbol{v^\prime} = (\beta/\alpha) \boldsymbol{v}$.  A common example is making the change $H_0^\prime = H_0/\alpha$ by changing the zero-point of the distance relation.  
Velocities are unchanged if $\beta$ = $\alpha$, so  $m^\prime = \alpha m$, ie $mH_0$ values are constant.  Another example is Kepler's third law of planetary orbits which, since Newton, has been explained by: if $\gamma = 1$, then $\beta^3 = \alpha^2$.

However, we consider that the zero-point of the distance scale is firm now, so we must keep distances unchanged, $\beta = 1$ and velocities are also unchanged, while considering changes in  both \Ho\ and $\Omega_m$; there is no simple scaling relation for that. 
For the change from $H_0=75$ to 70, $\alpha=1.054$ and velocities will drop by $1/\alpha$.  
If we go ahead anyway and use this $\alpha$ with $\beta=1$ to scale the paths in the best solution, we would need to reduced masses by $\alpha^2$.  
This will maintain the location of all turnaround spheres in the sample, since pairs with relative velocities of zero would remain at zero.  
Since these high density regions are not much affected by the orphan particle density or the external tidal field, any good new solution probably will have masses scaled by approximately this value; therefore this is a good place to start the search for a best solution.  

The external field $g_{ext}$ varies as $H_0\Omega_m^{0.4}$ (see Eq.~\ref{eq:gext} because it is based on the observed peculiar velocity field and, since we have $\Omega_m = 0.137/h^2$, it varies as $H_0^{0.2}$.  Hence, $g_{ext}$ drops only by 1.4\% as \Ho\ goes from 75 to 70, but by design, its integral over time should result in the about the same extra velocity field.

\begin{figure}
\epsscale{1.2}
\plotone{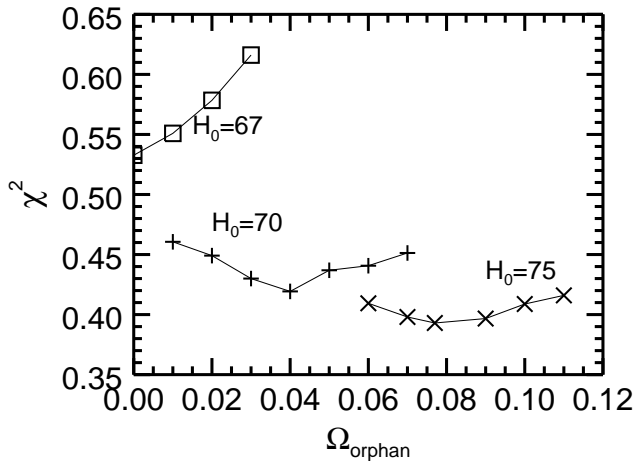}
\caption{\textbf{$\chi^2$ vs $\Omega_\mathrm{orphan}$} for best models
with $\Omega_\mathrm{orphan}$ and \Ho\ fixed. Squares, pluses, and crosses are for $H_0 = 67$, 70, and 75 \kmsMpc, respectively.  The other input parameters were varied to obtain minimum $\chi^2$ for each point. }\label{chiVos}
\end{figure}

After hunting for minimum $\chi^2$, a large reduction in $\Omega_\mathrm{orphan}$ and a further small decrease in $M_0$ or mass-to-light ratio is found (Fig.~\ref{chiVos}). 
This is understandable because 5.4\% drops in $v_m$ values become quite large at the edge of the sample, requiring an overall drop in mass.  The masses assigned to our targets can only be reduced a bit, consistent with the increased age of the universe, but the needed mass decrease can be achieved by dropping the smooth density component on scales where the overdensity is quite low.  

For \Ho\ = 67, the same rescaling procedure results in model velocities dropping by 9\%.  
This systematic with respect to observations cannot be compensated by the smooth component, because now $\Omega_\mathrm{orphan}$  has been pushed down to zero, and so that solution has a relatively higher $\chi^2$ value.  
We conclude that, if we adhere to the zero-point scale for distances which is consistent with $H_0 = 75$ out to 10,000 \kms, then $H_0 \lesssim 70$ is excluded from the dynamics of the Local Supercluster.
In any event, given that distances and observed velocities
are fixed by measurement, any choice of \Ho\ much below
75 \kmsMpc\ implies that we live in an under dense
location with a monopole outflow that is untenably large on the full scale of the \textit{CosmicFlows} sample \citep{2016AJ....152...50T},
and at odds with the linkage of the local type
Ia supernova scale to large distances \citep{Courtois2012a,2016ApJ...826...56R}.

\subsection{Virgo Cluster Vicinity}

In the best model, the Virgo Cluster redshift, $v_m = 1029$ \kms, matches well the value, $v_o=1030$ \kms\ from 97 2MASS galaxies with distance measures that place them in the cluster \citep{2017ApJ...843...16K}.
The distance to the cluster is 16.0 Mpc, so with \Ho\ = 75~\kmsMpc, the observed peculiar motion is $v_o - H_0d = -170$ \kms. About 45 \kms\ of this can be attributed to the external gravity field.

For groups within 15\degr\ of the cluster, where the peculiar velocities should be highly sensitive to the Virgo Cluster mass, the model seems to fit as well as expected, although several masses are somewhat deviant.   
However, near the Virgo Cluster, small distance errors can result in large velocity deviations. Conversely, distances can be accurately determined from redshifts, provided a reliable model.  
In Fig.~\ref{fig:DvsV15}, the run of velocity with distance is plotted with squares for the model and circles for the observations.  A vertical line connects the two.  
There were large discrepancies in velocities with NGC~4636 and NGC~4600, but we were able to adjust their distances within their $1\sigma$ uncertainties to obtain satisfactory velocities.  
Where a distance was adjusted, a horizontal line starts at the observed distance and ends at the displaced distance.

\subsection{Virgo turnaround and bound radii}

The region around an overdensity where expansion has been halted results in two additional points along the line-of-sight that manifest the same redshift as the overdensity  itself  \citep{Tonry1981}.
In the approximation of the collapse of a spherical mass distribution, the enclosed density divided by critical density of this ``first" turnaround point $r_{1t}$ (to differentiate it from the turnarounds occurring after one or more crossings) depends only on $\Omega_m$ (because it depends on $H_0t_0$ with $\Omega_\Lambda = 0$):
\begin{align}
\Omega_{1t} &= \frac{\rho_{1t}}{\rho_{crit}} = \bigg(\frac{2}{\pi} f(\Omega_m)\bigg)^{-2}\label{eq:o_tr1}\\
f(\Omega_m) &= (1- \Omega_m)^{-1} - \frac{\Omega_m}{2}(1-\Omega_m)^{-3/2}\eta\\
\eta &= \cosh^{-1}(2\Omega_m^{-1}-1).\label{eq:o_tr3}
\end{align}
For our fiducial case of $H_0=75$ and $\Omega_m=0.244$, then $\Omega_{1t} = 3.59$.  
In the case of the Virgo Cluster, there is sufficient measurement accuracy of distances and velocities to distinguish the turnaround surface about the cluster.  It is interesting to compare our best model value of the corresponding $\Omega_{1t}$ with radially symmetric collapse to learn how much non-radial motions affect this estimate.
By placing zero mass particles along the line to the Virgo Cluster, centered at supergalactic position $\boldsymbol{x} = [-3.6, 15.6,-0.7]$ Mpc, one develops the triple-valued velocity$-$distance relation, shown as the solid line in  Fig.~\ref{fig:DvsV15}.  
We can also search along the cardinal axes centered on the cluster to find where the radial component of the peculiar velocity cancels the Hubble expansion and the cluster would appear to have $v_m=0$ relative to the moving center of the cluster. 
Table~\ref{tbl_tr} gives information on these 6 turnaround points, including distances from the cluster, distances from us, (x,y,z) components of the locations, and components of the peculiar velocities in the sample frame of reference.  

\begin{deluxetable}{rrrrrrrrr}
\tablecolumns{9} 
\tabcolsep=0.11cm
\tablecaption{6 Points of Turnaround to Virgo Cluster\label{tbl_tr}} 
\tablehead{ 
\colhead{axis}    &  \dcolhead{D_{Vir}} &   \colhead{D}   & 
\colhead{SGX} & \colhead{SGY}   & \colhead{SGZ}    & \dcolhead{u_x} &
\dcolhead{u_y}    &  \dcolhead{u_z}\\
\nocolhead{one} &   \colhead{Mpc}   & \colhead{Mpc}  & \colhead{Mpc}  & \colhead{Mpc} & \colhead{Mpc} 
 &\colhead{km/s} & \colhead{km/s}  & \colhead{km/s}
}
\startdata 
+X & 6.50 & 15.9 &   2.9 & 15.6 & -0.7 & -724 &   14 &  291\\
-X & 6.75 & 18.7 & -10.4 & 15.6 & -0.7 &  269 &  296 &  -37\\
+Y & 7.40 & 23.3 &  -3.6 & 23.0 & -0.7 & -453 & -500 & -185\\
-Y & 7.05 &  9.3 &  -3.6 &  8.5 & -0.7 & -216 &  583 & -467\\
+Z & 7.95 & 17.6 &  -3.6 & 15.6 &  7.3 & -278 &  124 & -711\\
-Z & 7.85 & 18.1 &  -3.6 & 15.6 & -8.5 & -296 &   59 &  469\\
\enddata 
\end{deluxetable} 

\begin{figure}
\epsscale{1.2}
\plotone{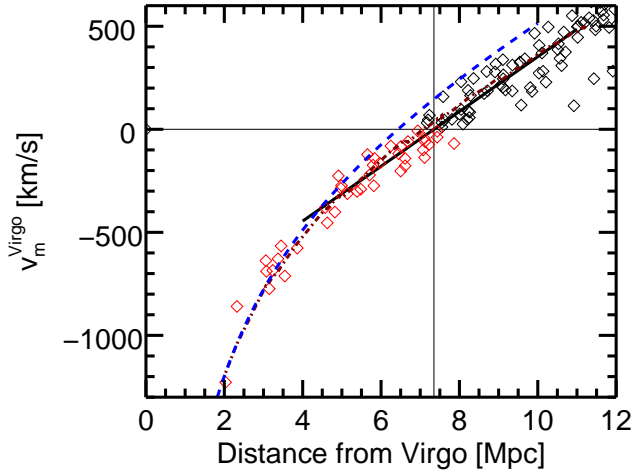}
\caption{Radial velocities in the Virgo Cluster frame $v_m^{\mathrm{Virgo}}$ versus distance from the cluster center in the best model.  The linear fit between 4 and 10 Mpc (solid black line) crosses zero velocity  at 7.35 Mpc.  Masses with negative radial velocities are associated with red squares, positive ones with black squares.  The spherical collapse model velocities are also shown as lines for just the Virgo Cluster mass (blue dashed) and for all mass internal to the radius (maroon dash-dotted).  The latter spherical model fits the points nearly as well as the linear fit.}\label{fig:rv_virgo}
\end{figure}  

The average of these turnaround radii is 7.25 Mpc, which, at the distance of the cluster subtends $\sim 25\fdg$9 in our sky.  
The range of turnaround points stretch from 6.5 to 7.95 Mpc.  

In Fig~\ref{fig:rv_virgo}, we plot radial velocities, calculated from peculiar velocities in our best model, in the Virgo Cluster frame of reference, $v_m^{\mathrm{Virgo}}$, versus distance from the cluster center.  
By fitting the points with a line from 4 to 10 Mpc, we find the turnaround point ($v_m^{\mathrm{Virgo}}=0$) is at $7.35 \pm 0.40$ Mpc.  
Weighted averaging the two estimates puts the turnaround point at $r_{1t}=7.3 \pm 0.3$ Mpc.  
The mass interior to this point including the Virgo Cluster and 40 nearby groups is $8.3 \pm 0.3 \times 10^{14} M_\sun$ and the range of $\Omega_{1t}$ is 3.1 to 3.7 ($\delta=11.7$ to 14.2).  
A turnaround mass of $6.6 \pm 0.9 \times 10^{14} M_\sun$ found by \cite{2016MNRAS.460.2015S} is in reasonable agreement even though a very different procedure was used.  
If we use the spherical infall model (Eqs.~\ref{eq:o_tr1}-\ref{eq:o_tr3}) result that $\delta_{1t}=3.6$, we would expect turnaround at 6.5 Mpc if only our Virgo Cluster mass is used, and 7.1 Mpc if we also include consistently the masses near the cluster.

\begin{figure}
\epsscale{1.2}
\plotone{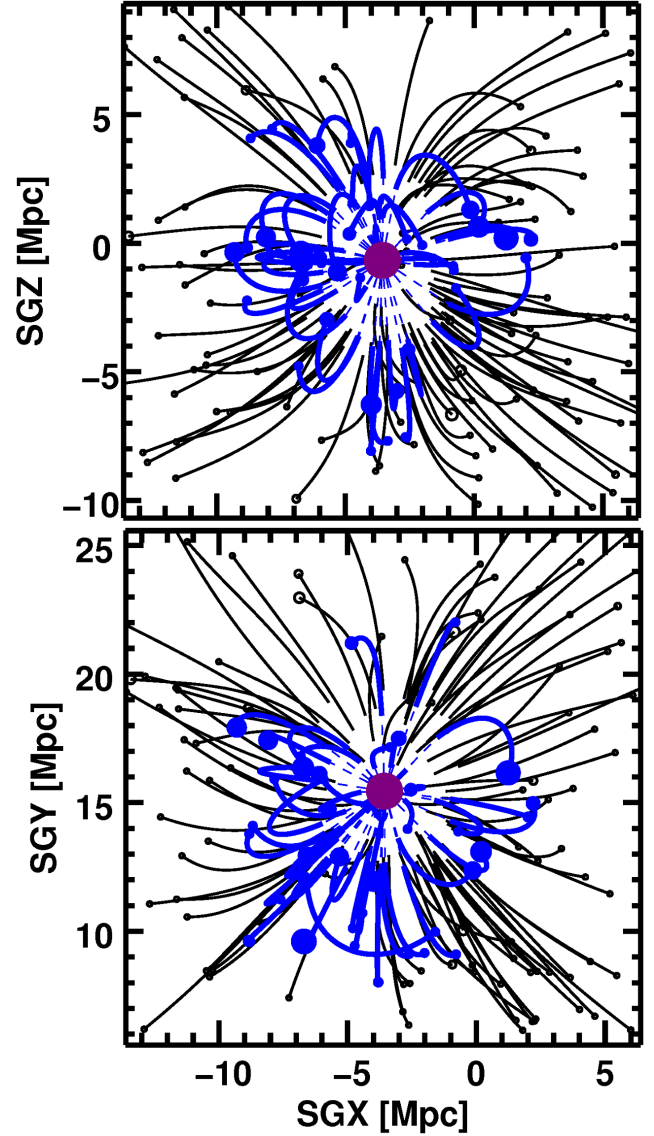}
\caption{\textbf{Paths of galaxies near the Virgo Cluster} in physical coordinates.  Paths with present day negative $v_m^{\mathrm{Virgo}}$ are blue. Virgo Cluster representation is purple.}\label{fig:orb_tr}
\end{figure}

Fig~\ref{fig:rv_virgo} also includes a dash-dotted maroon line for the velocities from the spherical collapse model using the model mass profile.  Evidently the radial velocities follow the collapse model very well.  This is so despite the fact that there are substantial transverse velocities in this region, as is evident in a plot of the paths of galaxies around the cluster (Fig~\ref{fig:orb_tr}).

\begin{figure}
\epsscale{1.2}
\plotone{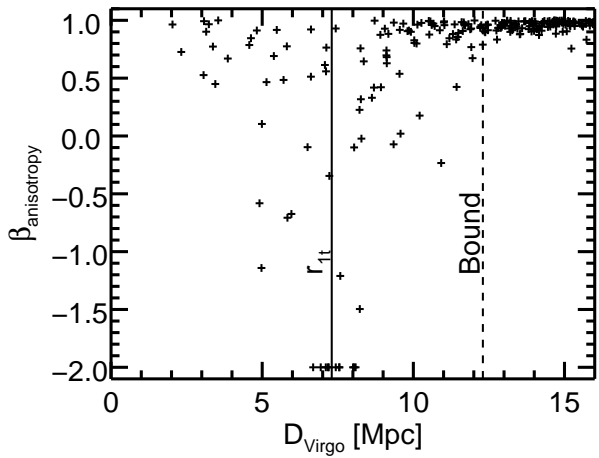}
\caption{\textbf{Anisotropy of physical velocities} in the Virgo Cluster frame of reference versus distance from the cluster center extracted from best model present day velocity results. The values are clipped to $-2$ if they are more negative than $-2$.
A vertical solid line is drawn at turnaround radius and a dashed line is drawn where enclosed density reached $\rho_{crit}$.}
\label{fig:beta}
\end{figure}

We examine the velocity anisotropy parameter $\beta = 1 - (v_\theta^2 + v_\phi^2)/(2v_r^2)$, where $v_\theta,v_\phi$, and $v_r$ are the two tangential and one radial physical velocities relative to the Virgo Cluster from the model.  If the velocities are more radial than anisotropic, then $\beta > 0$ and if they are more tangential, then $\beta < 0$.  
In Fig.~\ref{fig:beta}, a plot of $\beta$ versus distance from the cluster center, one sees that outside of 12.5 Mpc the motions are highly radial, but from 8 to 12.5 Mpc the velocities become more isotropic, and from 4 to 8 Mpc, near $r_{1t}$, there are many highly tangential orbits.  This is not too surprising  because the radial component of the velocity is highly diminished here.

\begin{figure}
\epsscale{1.2}
\plotone{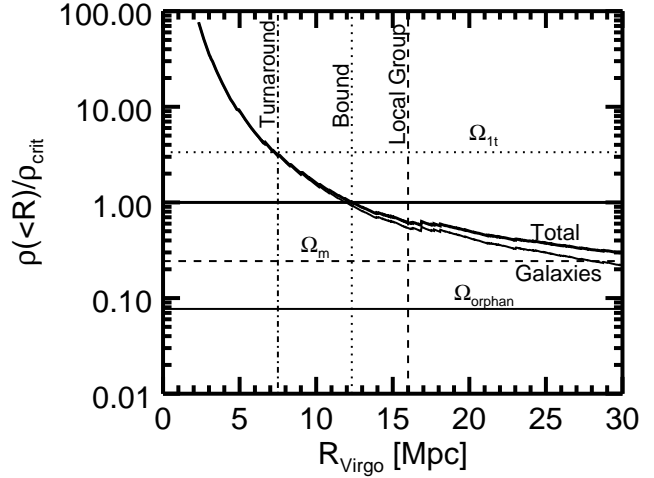}
\caption{\textbf{Enclosed density profile of the Local Supercluster} centered at Virgo Cluster for best model (marked Total).  
Density is divided by critical density.  Line marked Galaxies does not include the $\Omega_{\mathrm{orphan}}$ component.
Vertical lines are at Virgo turnaround (dash-dotted), the edge of the bounded region (dotted), and the Local Group distance (dashed).  The horizontal lines are at turnaround density $\Omega_{1t}$ for a spherical collapse model (dotted), mean matter density (dashed) and the smooth density component (solid).  Note the total density drops to the global mean value at the edge of our core region.
}\label{fig:density}
\end{figure}

We can use the mass versus light and $\Omega_\mathrm{orphan}$ determinations to calculate the spherically averaged internal density profile, relative to critical density, centered on the Virgo Cluster (Fig.~\ref{fig:density}).  
The gravitationally bound radius, where the density drops to $\rho_{crit}$ occurs at 12.3 Mpc from the cluster.
We live only 3.7 Mpc beyond that.  
The non-zero cosmological constant moves this shell slightly inward, but the planar structure of the Local Supercluster moves it slightly outward for most galaxies at the boundary. Finally, near the edge of the core region, at about 30 Mpc, the enclosed density has dropped to the global mean density.  For the best $H_0=70$ run, the profile of $\rho/\rho_{crit}$ is nearly identical because both the density and critical density drop by the square of the \Ho\ change; however, since $\Omega_m$ is a bit higher, the outer radii of the sample are slightly under dense.

\subsection{Collapse to the Supergalactic Equatorial plane}
\label{sec:collapse}

The most prominent feature of the flow field within 38 Mpc seen in Fig.~\ref{fig:orbitsxyzcz}a,b is the 
downward motion toward the supergalactic equator of all galaxies above the equatorial 
plane (SGZ $>$ 0) across the entire region.  
This phenomenon is most spectacularly revealed in the animated and interactive figures (see Fig.~\ref{fig:animation}).
Orbits curl left and down in SGY$-$SGZ toward the Fornax Cluster at negative SGY and curl right in 
the same projection toward the Virgo Cluster at positive SGY.    
The overwhelming downward flow is a manifestation of the general emptiness of the entire nearby 
region above the equatorial plane.  
This extension of the Local Void across to a void on the far side of the Virgo Cluster, the Virgo Void, 
is captured in Fig.~21 by \citet{Courtois2013}.

Motions below the equatorial plane tend to be upward but, except in a few areas, the trend is modest.  There are more substantial structures at negative SGZ.

\subsection{Leo Spur Anomaly\footnote{In the ensuing sub-sections the names of filamentary structures are drawn from the terminology introduced in the Nearby Galaxies Atlas and Catalog \citep{1987ang..book.....T,1988ngc..book.....T}.}}

\begin{figure}
\epsscale{1.2}
\plotone{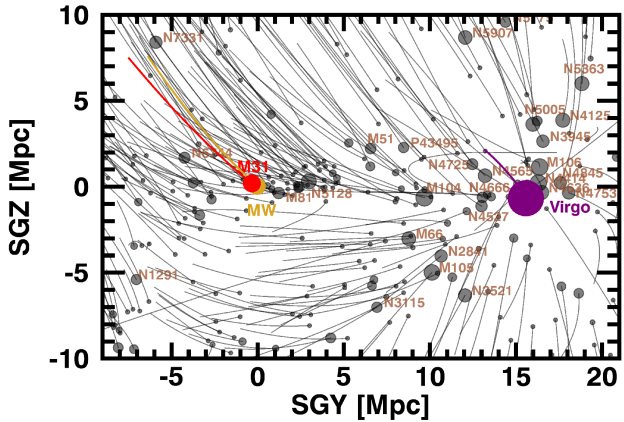}
\caption{\textbf{Paths in the region from the Local Group to Virgo} 
(zoom in on Fig.~\ref{fig:orbitsxyzcz}b). 
The region includes the Local Sheet and Virgo Cluster in the SGZ=0 equatorial plane, 
the Local Void at positive SGZ where flows are overwhelmingly downward, 
and structures at negative SGZ that are being attracted toward the Virgo Cluster.  
The MW and M31 orbits are yellow and red and the Virgo orbit is purple. 
Depth of the plot is $-10 < {\rm SGX} < +10$ Mpc.}
\label{fig:orbitszoomyz}
\end{figure}

It has been noted that galaxies in the Leo Spur and other galaxies below the supergalactic equator have 
highly negative peculiar motions \citep{Tully2008, 2015ApJ...805..144K}.
Fig.~\ref{fig:orbitszoomyz} shows the galaxy paths  between the Local Group region and the Virgo Cluster.  
While the galaxies in and near the supergalactic plane including the MW are rushing at high speed out of the Local Void toward $-$SGZ and toward the Virgo Cluster at +SGY \citep{2017ApJ...835...78R},  below the plane the flow pattern is a gentle arch headed to $-$SGZ at negative SGY values and mildly headed to +SGZ as one approached the Virgo Cluster.
The mass within the supergalactic plane shields the region below the supergalactic plane from the Local Void 
repulsion, plus any residual forces are balanced by another void at $-20 < SGZ < -10$ (Fig.~\ref{fig:orbitsxyzcz}b).
The overall effect is that peculiar velocities are nearly perpendicular to our line of sight everywhere just below the plane.  
The negative peculiar motions of galaxies below
the equatorial plane mainly reflect the MW motion toward negative SGZ.

\subsection{Leo Cloud}

\begin{figure}
\epsscale{1.1}
\plotone{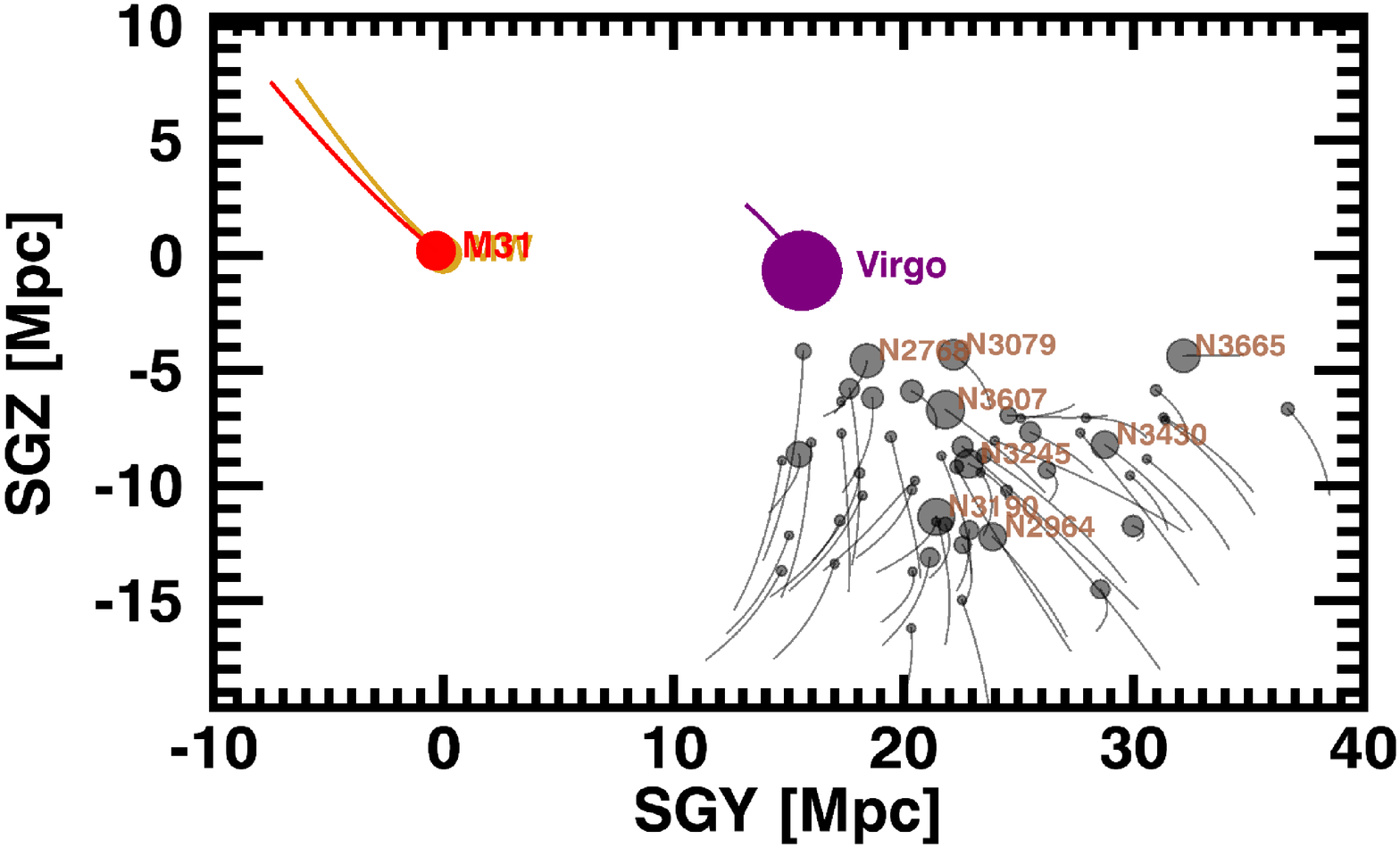}
\caption{\textbf{Orbits in the region of the Leo Cloud.}  The proximity of the Virgo Cluster is 
shown (in purple) as well as the displacement of the MW and M31 (yellow and red) toward the 
	Leo Cloud.  The depth of the plot is $-5 < \rm{SGX} < +29$ Mpc.}
\label{fig:leocloudyz}
\end{figure}

Galaxies in the structure called the Leo Cloud have peculiar motions with respect to the 
MW of order $-700$~\kms, 
the most extreme deviations found across an extended region anywhere within the 38 Mpc volume.
The projected SGY$-$SGZ location of the Leo Cloud is seen in Fig.~\ref{fig:leocloudyz}.
The cloud is found immediately behind the Virgo Cluster and is experiencing a substantial coherent flow toward the cluster.
The red and yellow orbits in the figure show the model trajectories of M31 and the MW.
It is seen that our peculiar velocity and the bulk Leo Cloud peculiar velocity are closely anti-aligned.

The substantial flows within the local region can lead to considerable projection confusion if independent distances are not available.
The projection of the Leo Cloud onto the Leo Spur is a case in point.
Near SGL$\sim95^{\circ}$, SGB$\sim-20^{\circ}$, galaxies in the vicinities of the Leo Spur foreground NGC 3379 
and NGC 3623/27/28 groups at 10 Mpc and the Leo Cloud background NGC 3607 and NGC 3457 groups at 22 Mpc have similar, 
slightly overlapping observed velocities \citep{2017ApJ...843...16K}.

\subsection{Structure Immediately Below the Equatorial Plane}

Attention was already drawn to the occurrence of filaments immediately below the supergalactic equator and at positive SGY by  \citet{1982ApJ...257..389T}.  The structures include the Leo Spur, Leo Cloud, Ursa Major Southern Spur, Crater Cloud, and the Antlia$-$Dorado filament.  These structures that lie as close to the plane as SGZ $\sim -2$ Mpc, extending down to $\sim -10$ Mpc, are {\it not} participating in the downward flow described in sub-section~\ref{sec:collapse}.  The transition in deviant velocities toward negative SGZ at the equatorial plane is abrupt.

\subsection{Fornax Region}

\begin{figure}
\includegraphics[width=\columnwidth,clip=true]{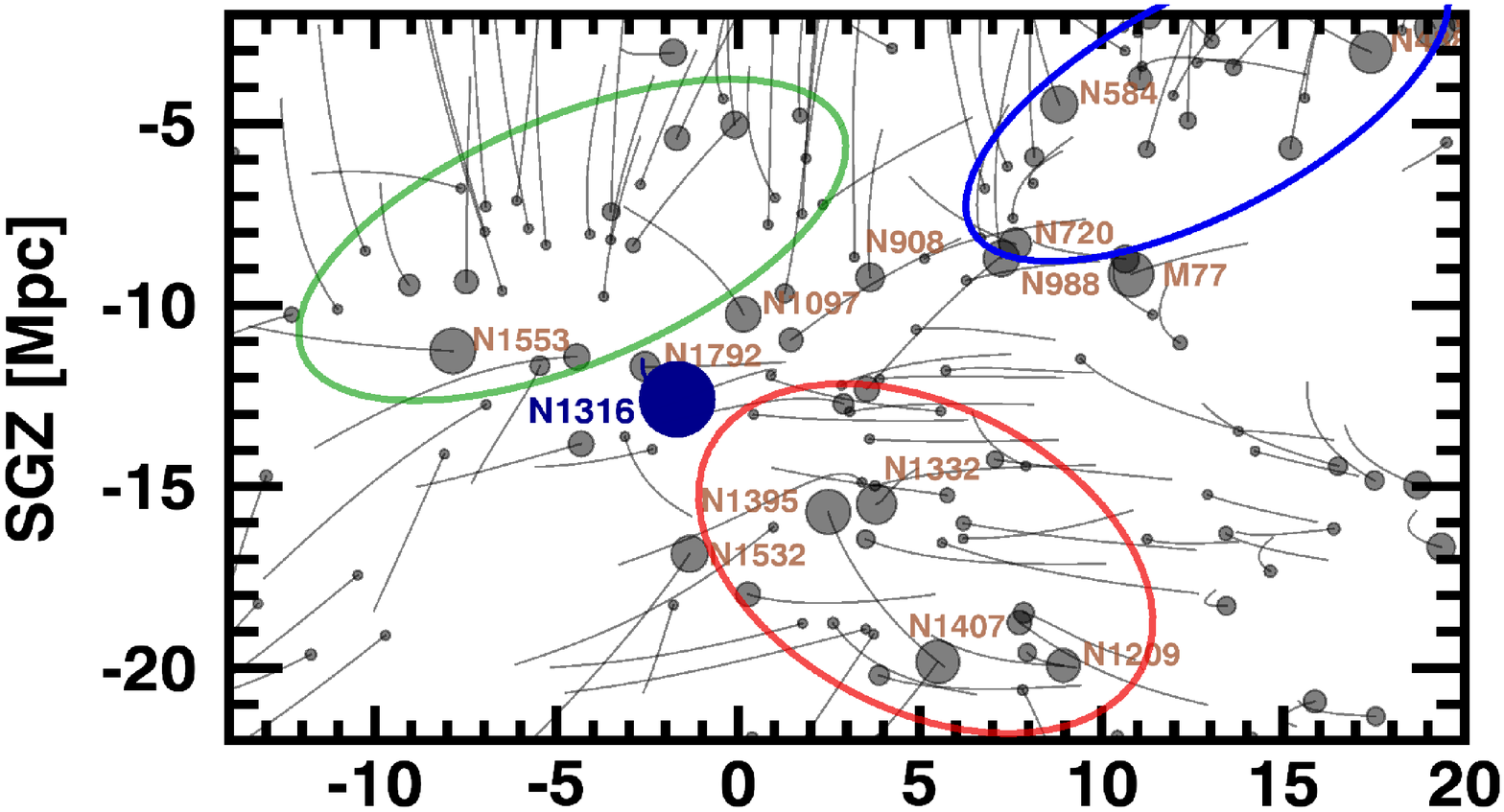}
\vskip -.45cm
\includegraphics[width=\columnwidth,clip=true]{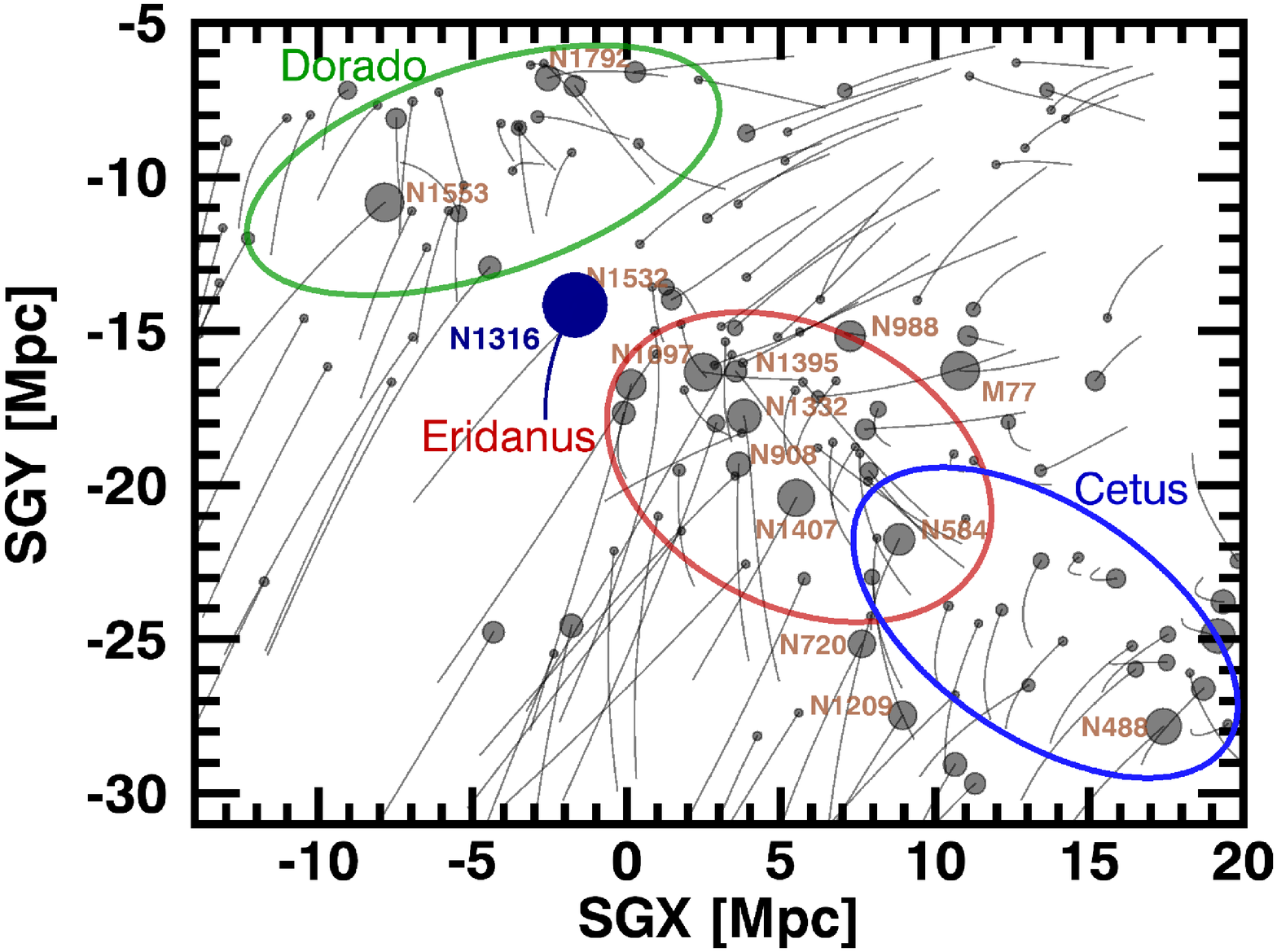} 
\caption{\textbf{Two projections of orbits in the region of the Fornax Cluster.}  The location of the Fornax Cluster is indicated by the filled blue circle.  
The Eridanus, Cetus-Aries, and Dorado clouds are enclosed in red, blue, and green ellipses. In the SGX$-$SGY projection the Cetus$-$Aries Cloud projects directly onto the Eridanus Cloud.  
Dominant flows are the motions out of the Local Void to negative SGZ at the top of the top panel and the motions out of the Sculptor Void toward positive SGY at the bottom of the bottom panel.}
\label{fig:fornax}
\end{figure}

The confinement of galaxies to a thin supergalactic equatorial plane dissipates beyond 10 Mpc south of the MW plane, 
although the general velocity trends seen north of the MW are maintained.  
This southern volume can still usefully be separated between structures above and below the 
supergalactic equator.  
Of primary interest below the equator are the Fornax Cluster and closely related Eridanus Cloud, 
the Cetus$-$Aries Cloud, and the Dorado Cloud.  
The region is illustrated with SGX$-$SGY and SGX$-$SGZ projections in Fig.~\ref{fig:fornax}.  

The two major flow features in the Fornax region can both be ascribed to repulsion from voids.
The Local Void expansion manifested in downward SGZ motions reaches the Cetus-Aries and Dorado clouds at SGZ roughly $-4$ to $-8$ Mpc.
The Fornax$-$Eridanus structures appear to be isolated from the Local Void expansion by their locations directly under Cetus$-$Aries in SGZ.
Meanwhile, the pronounced flow toward positive SGY experienced across the region can in large measure 
be attributed to the expansion of the Sculptor Void, a large under density at negative SGY that is 
bounded on the back side by the Southern Wall at $\sim 80$~Mpc \citep{1990AJ.....99..751P,1998lssu.conf.....F}.

The Dorado Cloud is a part of a filament that extends through the MW to become the Antlia Cloud north of the Galactic plane.  The filament is marked as it passes through the zone of obscuration by foreground Puppis groups  \citep{1992A&A...266..150K,2016AJ....151...52S}.  The entire region is participating in a flow toward the Centaurus Cluster.

\subsection{South Galactic, North Supergalactic}

\begin{figure}
\includegraphics[width=\columnwidth,clip=true]{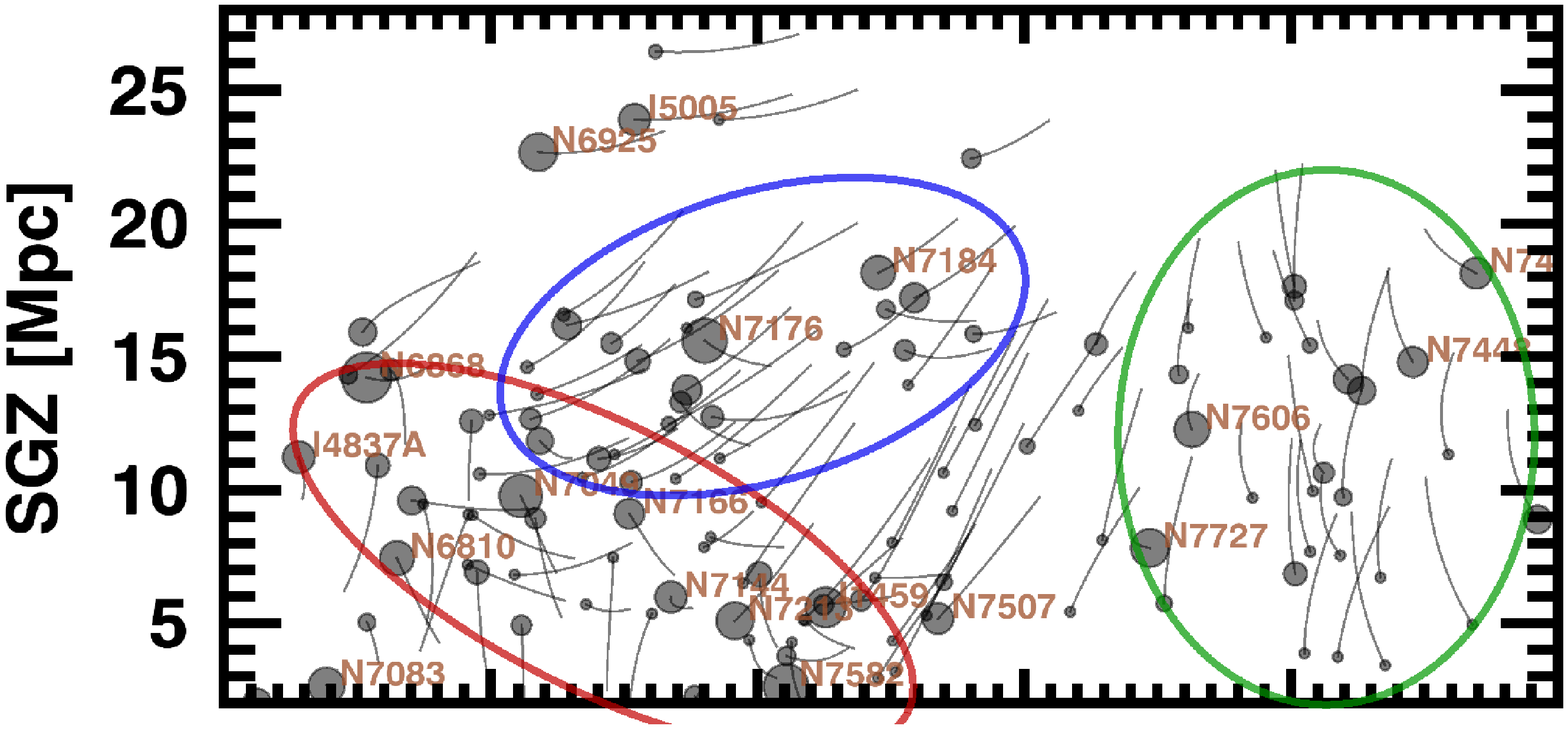}
\includegraphics[width=\columnwidth,clip=true]{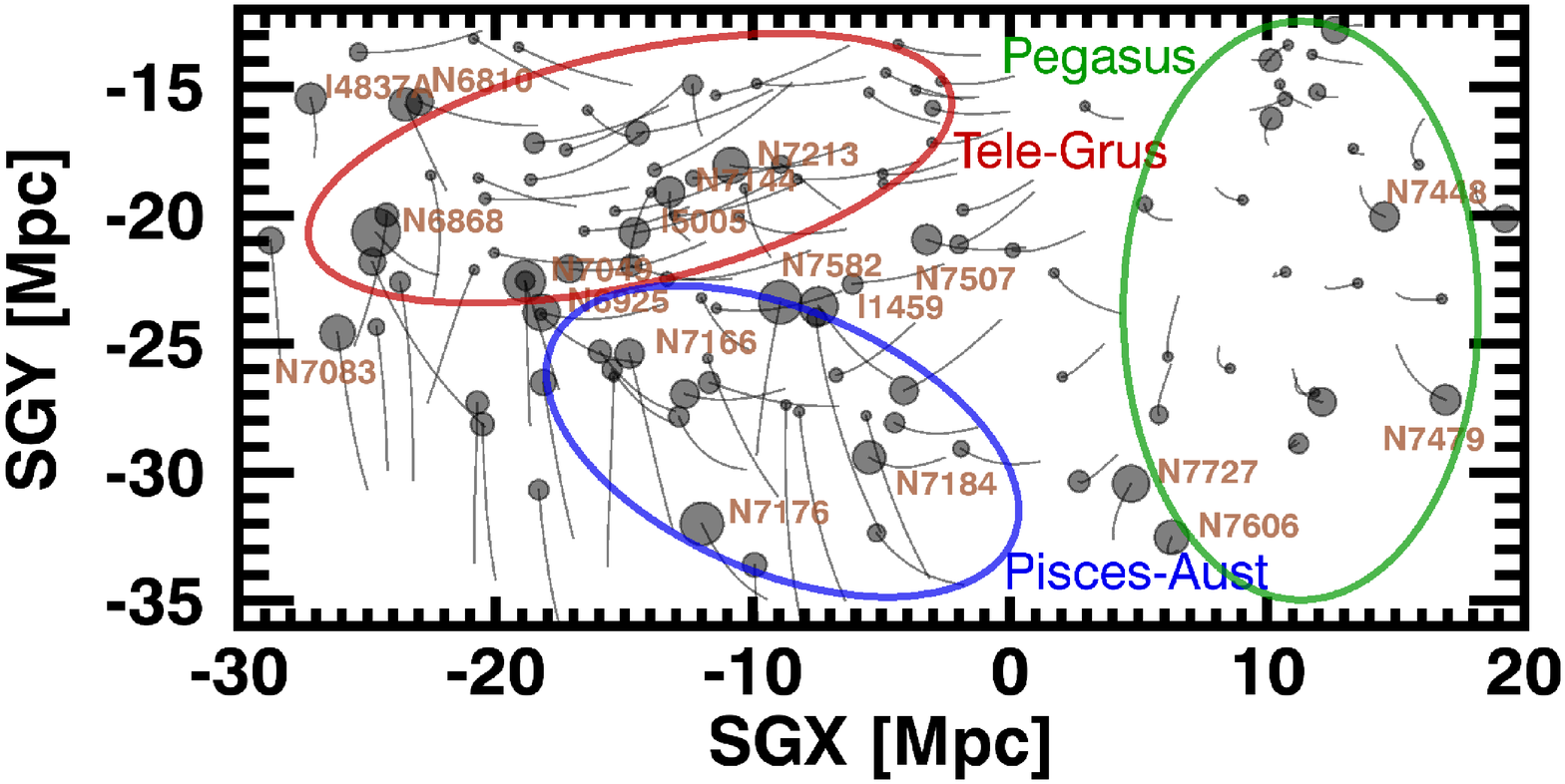}
\caption{\textbf{Two projections of orbits around the Telescopium$-$Grus, Pisces-Austrinus, and Pegasus clouds.}  These entities are enclosed in red, blue, and green ellipses, respectively.  The former two clouds are participating in the flow toward the Great Attractor while the latter is strongly influenced by the Perseus-Pisces complex.}
\label{fig:telgrus}
\end{figure}

Within the 38 Mpc volume, only here (as seen in Fig.~\ref{fig:telgrus}) 
are there extended structures with positive peculiar motions with respect to the MW, reaching $\sim 300$~\kms\ in the Pegasus Cloud.   
The other principal structures are the Telescopium$-$Grus Cloud and the Pisces$-$Austrinus Spur.  
Motions are downward in SGZ across the region in response to the emptiness of the supergalactic north.  
The other pronounced systematic is a divergence in the velocity field around SGX$\sim+4$~Mpc.  
At more negative SGX, galaxies are participating in the large scale flow toward the Centaurus/Great Attractor region, 
while at more positive SGX this motion is stalling due to the influence of the major Perseus$-$Pisces filament, 
a phenomenon discussed further in sub-section~\ref{sec:ga}.

\subsection{The Centaurus-Puppis-PP Filament}

Evidence is emerging that a substantial filament emanating from the Centaurus Cluster runs all the way to the region of the important 
Perseus$-$Pisces filament  \citep{Haynes1988}, passing behind the plane of our galaxy near the south supergalactic pole at SGZ$\sim-25$~ Mpc.  
Because parts of its path are obscured, pieces have acquired different names.  
The well defined initial portion in the north galactic hemisphere near the Centaurus Cluster was called the 
Antlia$-$Hydra Cloud in the Nearby Galaxies Atlas \citep{1987ang..book.....T} and the Hydra Wall by \citet{1998lssu.conf.....F}.  
The heavily obscured leg near the MW was called the Lepus Cloud in the Nearby Galaxies Atlas and the Puppis filament by \citet{1992A&A...266..150K}.  
HI surveys of the zone of obscuration have confirmed the substantial nature of the structure \citep{2016AJ....151...52S}.  
The term Antlia Strand was introduced by \citet{Courtois2013} in the context of five apparent strands feeding into the Centaurus Cluster.  
The name used here was given by \citet{2017ApJ...845...55P}.
Their Wiener Filter reconstruction of structure related to the discussion in Section~\ref{sec:external} reveals the large scale continuity of the feature across $\sim130$~Mpc from the Centaurus Cluster to the Perseus$-$Pisces filament.

\begin{figure}
\epsscale{1.25}
\plotone{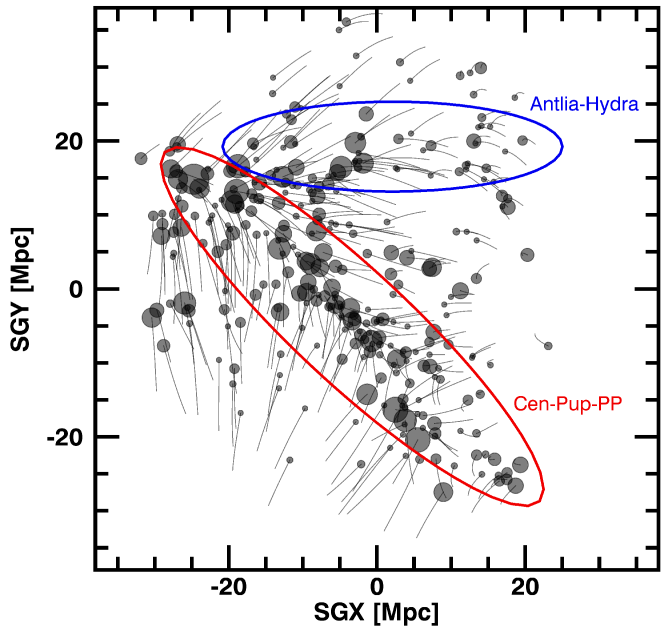}
\caption{\textbf{The Centaurus-Puppis-PP Filament and Antlia$-$Hydra Cloud.} Orbits of masses at SGZ $< -14$ Mpc. These two structures participate in a coherent flow toward the Centaurus Cluster.}
\label{fig:antliaxy}
\end{figure}

The coherent flow within this structure is seen in Fig.~\ref{fig:antliaxy}.  Across almost the entire extent, the flow is toward the location of the Antlia Cluster and from there toward the Centaurus Cluster.  However, around SGX$\sim10$, SGY$\sim-20$~Mpc the flow stalls as the filament passes beyond our 38 Mpc boundary.  The Wiener Filter model reveals that the flow is diverted toward the Perseus$-$Pisces filament at greater distances.

\subsection{Motion toward the Great Attractor}
\label{sec:ga}

The dominant features of the velocity field in the volume under consideration are the downward flow 
everywhere above the supergalactic equator and the flow toward the edge of the study region around 
SGX$\sim-40$, SGY$\sim20$~Mpc, near the Centaurus Cluster and the putative 
Great Attractor \citep{Shaya1984d, 1987ApJ...313L..37D, Lynden-Bell1988}.
It is to be reminded that the rest frame in the present discussion is with respect to the center of mass of the ensemble involved in the orbit reconstruction.
This reference frame has a bulk motion with respect to the frame established by the Wiener Filter model (Section~\ref{sec:external}) of $v_{SGX},v_{SGY},v_{SGZ} = [-212, 95, -106]$~\kms.
Hence the locally derived flow  in most places within 38 Mpc toward negative SGX is an additive part of a coherent flow in that direction.

While this flow toward Centaurus/Great Attractor is pervasive, attention has been drawn to regions at the positive SGX edges of the study region where this flow stalls and there are hints of an oppositely directed flow.  Consideration of the Wiener Filter model on larger scales reveals that the positive SGX edge of the 38 Mpc volume approximates the boundary between the Laniakea Supercluster \citep{Tully2014a}, our home basin of gravitational attraction, and the adjacent Perseus$-$Pisces and Arrowhead structures \citep{Haynes1988,2015ApJ...812...17P}.

\section{Discussion}
The best model has a median reduced $\chi^2$ of $\sim 0.4$ which indicates that the flow is modeled well even when a constant mass-to-light ratio is used. 
A slight improvement in $\chi^2$ was found with a mass-to-light ratio that varied with luminosity.  
The average value of the reduced $\chi^2$ is $\sim 1.5$ which, given that only simple orbits were accepted, implies that only a few of the galaxies or groups have executed complex paths of interactions.  
Some experimentation has shown that often, when a group has a highly deviant model velocity, a complex interaction with a neighbor can be found that greatly improves the agreement, but not always.  In particular, the Virgo Cluster had only one solution path for a given set of input parameters, despite much effort to force it into other paths.  

It is generally assumed that the distribution of galaxies is biased with respect to the underlying matter distribution.  The dependence with a power law in luminosity for the mass-to-light ratio, plus a uniform unclustered background density in $\Omega_{\mathrm{orphan}}$, that we interpret as matter that has neither fallen into discrete objects or has escaped them, hopefully provide good proxies for the biasing phenomenon. 
The density of this intergalactic medium in the model is $\Omega_\mathrm{orphan} = 0.077$, which is $\sim30$\% of the total mass density and implies a primarily unobserved intergalactic medium is a very important component of the universe.  
Of course, our model is simplistic.  
Associating substantial residual mass with an intergalactic reservoir is plausible but it is naive to consider it to be of uniform density.  Comparisons with simulations are necessary. 
This parameter is rarely discussed in N-body simulation results, but it is in line with the wide range of values claimed in the few studies that have mentioned it \citep{Angulo&White2010, Libeskind_etal2012}.  Unfortunately, NAM does not provide insight into details in the distribution of orphan particles or its history.
  
The summed model mass of M31 and MW (the only galaxies included in the representation of the Local Group mass) is $5.15 \pm 0.35 \times 10^{12} M_\sun$. 
This mass is in line with the simple radial collapse timing argument \citep{1959ApJ...130..705K} which gives a mass of $4.9 \times 10^{12} M_\sun$ with our values of separation and approach speed.
Solutions were found in which M31 moves along the plane while the MW is just now falling into the plane, as it does in our preferred solution. 
This family of solutions has total Local Group masses of up to $\sim 7 \times 10^{12} M_\sun$ and transverse velocities of $\sim 130$ \kms\ for M31.
The family of solutions in which the MW moves along the local plane and M31 starts at high SGZ have poorer $\chi^2$ values because the SGZ component of velocity for the MW is too low.  
The best solutions were those in which the proper motion of M31 is $pm < 25~\mu\rm{as~yr}^{-1}$.  
At these value, the MW and M31 are headed almost directly toward each other, and the low $\chi^2$ results indicate that this is the correct solution for the MW motion relative to the nearby galaxies.  
 
\section{Conclusions}

The availability of many hundreds of distances to galaxies permits a detailed study of galaxy flows in the nearby universe where dynamics are in the non-linear regime due to significant density concentrations.   
Numerical Action Methods provide a powerful tool for the reconstruction of orbits of the major mass elements and of tracer elements that constrain the model with knowledge of distances.  We have chosen to construct paths of galaxies from early time until the present in a region 38 Mpc about us which encompasses the Local Supercluster.  
The infall toward the Virgo Cluster, the dominant cluster in the Local Supercluster, is well documented, with the expected 'triple value region'  where the same radial velocity appears at three nearby distances along the line of sight. 
The infall pattern demands a total mass associated to the Virgo Cluster of $6.3 \pm 0.8 \times 10^{14} M_\sun$. The cluster is the most substantial collapsed structure in the sample, with an infall domain that extends on average to $r_{1t}=7.3\pm0.3$ Mpc. 
All galaxies within 12.3 Mpc, where the enclosed density drops to $\rho_{crit}$, will eventually fall into the Virgo Cluster.
We are saved from this fate by a margin of only 3.7 Mpc.  
At the Local Group distance from the Virgo Cluster, the enclosed Virgo centered matter density is  $0.61 \rho_{crit}$ in both the $H_0=75$ \kmsMpc\ and $H_0=70$ \kmsMpc\ cases (with all distances and $\Omega_m h^2$ fixed), but in terms of overdensity, $\delta = 1.5$ in the  $H_0=75$ case and $\delta = 1.2$ in the $H_0=70$ case.

We achieve a satisfactory $\chi^2$ fit with alternatively a constant mass-to-light assumption $M/L_{K_s} = 58 \pm 3~M_\sun/L_\sun$  or with the assumption of a mildly increasing mass-to-light ratio motivated by literature sources $M/L_{K_s} = 40 \pm 2~L_{10}^{0.15} M_\sun/L_\sun$ ($L_{10}$ is $K$-band luminosity in units of $10^{10}~L_\sun$), which would put the Virgo Cluster $M/L_K$ at $103 ~M_\sun/L_\sun$, whereas the model's best value is $M/L_K = 113 \pm 15~M_\sun/L_\sun$).  The mass of the Local Group, given by the sum of the masses of the MW and M31 is $5.15\pm0.35 \times 10^{12}~M_\sun$, which agrees very well with an updated timing-argument calculation.  The MW moves with the correct speed and direction to provide a proper reflection of its own velocity in measures of radial velocities of nearby galaxies when the MW and M31 motions are almost directly toward one another. 

An additional roughly smooth intergalactic medium was necessary to provide sufficient total density.  
The best value of $\Omega_\mathrm{orphan} = 0.077\pm0.016$ brings the mean density of the entire 38 Mpc core region to the global mean value.  But for the $H_0 = 70$ \kmsMpc\ case, then $\Omega_\mathrm{orphan} \sim 0.04$ and masses are down by roughly a factor of inverse square of the $t_0$ ages.

Besides Virgocentric flow, there are several other unambiguous flow patterns within the study region of 38 Mpc.  One of these is the now familiar flow toward the direction of the Centaurus Cluster and surrounding Great Attractor region, the core of the Laniakea Supercluster.  The apex of the dipole in the cosmic microwave background temperature map is in the same direction.  This flow dominates almost our entire region of study with the exception of regions at the one edge with most positive values of supergalactic SGX where  the attraction of the Perseus-Pisces complex is becoming competitive.  We draw particular attention to the Centaurus-Puppis-PP filament which is a coherent structure extending across the breadth of our volume but which has been poorly known because it is relatively far away and obscured over part of its length.  This filament is an important connector between the Laniakea and Perseus-Pisces basins of attraction. 

Perhaps the most remarkable feature for the combination of its pervasiveness and lack of familiarity is the downward flow of all galaxies with positive supergalactic SGZ across the entire region. 
The Local Void was already a known feature, causing evacuation flows bounded by the Local Sheet.  
It is now clearly revealed that the entire north supergalactic cap is under dense throughout the volume extending to 38 Mpc.  
The underdensity wraps around behind the Virgo Cluster, accounting for its motion toward us. 

Our modeling demonstrated to us the importance of including tidal influences on scales larger than we could consider in the construction of numerical action orbits.  
The very asymmetric distribution of mass concentrations beyond the Local Supercluster are important but equally so is the absence of mass in voids that impinge nearby but extend well beyond the core.  
The expansion of voids create coherent streaming on large scales and velocity discontinuities where matter piles up in walls.

\acknowledgments

We thank Jim Peebles for providing us with an early release of the NNAM code, and we have benefited tremendously from discussions with him. YH has been supported by the Israel Science Foundation (1013/12).  
The important TRGB component of distances came from Hubble Space Telescope observations and support for programs HST-GO-14636, HST-GO-14636, HST-GO-13442, HST-GO-12878, and HST-GO-12546 were 
provided by NASA through a grant from the Space Telescope Science Institute, which is operated by the Association of Universities for Research in Astronomy, Inc., under NASA contract NAS 5-26555.
\newpage

\bibliography{mytex16}
\bibliographystyle{apj}
\end{document}